\documentclass[12pt,a4paper]{article}
\makeatletter
\@addtoreset{equation}{section}
\makeatother

\usepackage[pdftex]{graphicx}
\usepackage{subfigure}
\usepackage{array}
\usepackage{amsmath}

\usepackage{rotating}
\usepackage{longtable}
\usepackage{mathtools}
\usepackage[numbers]{natbib}

\usepackage{array}
\usepackage{geometry}
\usepackage{color}
\usepackage{lipsum}

\usepackage{diagbox} 
\usepackage{pdfpages}
\usepackage{multicol}
\usepackage[utf8]{inputenc}
\usepackage[english]{babel}
\usepackage{mathtools}
\usepackage{bbm}
\usepackage{amssymb}
\usepackage{amsthm}
\usepackage{hyperref}
\usepackage{bm}
\usepackage{mathtools}
\usepackage{fancybox}
\usepackage{multirow}
\usepackage{caption}
\usepackage{subfigure}
\usepackage{algorithm}
\usepackage[affil-it]{authblk}
\usepackage[noend]{algpseudocode}
\usepackage{setspace}

\makeatletter
\newcommand*{\rom}[1]{\expandafter\@slowromancap\romannumeral #1@}
  
\makeatletter
\def\BState{\State\hskip-\ALG@thistlm}
\makeatother
\allowdisplaybreaks[4]

\usepackage{xcolor}
\renewcommand{\raggedright}{\leftskip=0pt \rightskip=0pt plus 0cm}
\title{{\bf Coefficient Shape Alignment\\ in Multivariate Functional Regression}}
\author[]{Shuhao Jiao\thanks{shuhao.jiao@cityu.edu.hk} }
\author[]{Ngai Hang Chan}
\affil[]{Department of Biostatistics,\\City University of Hong Kong, Hong Kong}

\date{}
\begin{document}
	\maketitle			
	\setlength\parindent{0pt}
	\setlength{\parskip}{1em}
	\theoremstyle{definition}
	\newtheorem{theorem}{Theorem}
	\newtheorem{ass}{Assumption}
	\newtheorem{lemma}{Lemma}
	\newtheorem{remark}{Remark}
	\newtheorem{prop}{Proposition}
	\newtheorem{Definition}{Definition}
	\newtheorem{cor}{Corollary}
\begin{abstract}
In multivariate functional data analysis, different functional covariates often exhibit homogeneity. The covariates with pronounced homogeneity can be analyzed jointly within the same group, offering a parsimonious approach to modeling multivariate functional data. In this paper, a novel grouped multiple functional regression model with a new regularization approach termed {\it ``coefficient shape alignment"} is developed to tackle functional covariates homogeneity. The modeling procedure includes two  steps: first aggregate covariates into disjoint groups using the new regularization approach; then the grouped multiple functional regression model is established based on the detected grouping structure. In this grouped model, the coefficient functions of covariates in the same group share the same shape, invariant to scaling. The new regularization approach works by penalizing differences in the shape of the coefficients. We establish conditions under which the true grouping structure can be accurately identified and derive the asymptotic properties of the model estimates. Extensive simulation studies are conducted to assess the finite-sample performance of the proposed methods. The practical applicability of the model is demonstrated through real data analysis in the context of sugar quality evaluation. This work offers a novel framework for analyzing the homogeneity of functional covariates and constructing parsimonious models for multivariate functional data.\\

\end{abstract}	
\newpage
\section{Introduction}
Functional data analysis (FDA) is an area of statistics that aims at modeling complex objects such as functions, images, shapes, and manifolds (see \cite{ramsay2005functional} and  \cite{wang2016functional}). Multivariate FDA, as a natural extension of FDA, involves increased complexity as it must address not only the infinite-dimensionality of each functional component but also the associations between different components within this infinite-dimensional space. In the realm of  regression involving multiple covariates, understanding these associations becomes crucial for accurately modeling the relationships between the dependent variable and the covariates. In principle, associated covariates in the same group typically share some kind of homogeneity, which forms the foundation for constructing the grouping structure. The homogeneity structure offers valuable insights into the connectivity or association between different covariates, making it crucial in multivariate FDA.  However, the challenge is that it is generally unknown which covariates should be grouped together. This uncertainty necessitates the development of methods to detect the underlying grouping structure, ensuring that the analysis effectively capture the relationships and interactions among functional covariates.

Regularization has become a popular tool for model selection and simplification (see e.g., \cite{tibshirani1996regression}, \cite{fan2001variable}, \cite{tibshirani2005sparsity}, \cite{yuan2006model}, \cite{james2009functional},  \cite{shen2010grouping}, \cite{zhang2010nearly}, \cite{zhao2012wavelet}, \cite{lian2013shrinkage}, \cite{ke2015homogeneity}, \cite{lin2017locally}, \cite{ma2017concave}, \cite{yeh2023variable}). Grouping pursuit, a special class of regularization, has been developed to cluster covariates or samples into groups based on the homogeneity of coefficients. 
However, existing literature on grouping pursuit mainly addresses regression with scalar or vector covariates and assumes coefficient equality. For example, fused LASSO (\cite{tibshirani2005sparsity}) seeks to fuse adjacent equal coefficients. \cite{shen2010grouping} and  \cite{ke2015homogeneity} study covariate clustering based on the equality of associated coefficients in linear regression. \cite{ma2017concave} focus on intercept fusion, aiming to identify the homogeneous groups of equal intercepts to cluster samples. More recently, \cite{she2022supervised} studied the grouping problem of multivariate features in a regression model, where each feature is a vector-valued covariate. However, the grouping structure considered therein is still based on coefficient equality. The infinite-dimensional nature of functional data introduces complexities that scalar-based approaches cannot adequately address, and thus in functional regressions, the concept of within-group homogeneity should encompass more nuanced forms of similarity beyond mere equality of coefficients. In addition, a major limitation of equal coefficient fusion is that it is {\it not} robust to covariate scaling because covariate scaling changes the difference of the associated coefficients. This issue motivates us to develop a new means to inclusively and robustly capture the homogeneity of different covariates. 
To overcome the limitations of existing methods, we develop a novel grouping structure and detection procedure that focus on coefficient shape commonality across multiple functional covariates. This new approach is designed to be more flexible and comprehensive, integrating equal coefficient fusion as a special case but extending its applicability to a wider range of scenarios. In addition, the grouping structure is invariant to covariate scaling. This property is crucial in grouping pursuit, as the variations of different covariates can vary significantly. Covariates with high data variation overshadow those with lower variation. To mitigate this undesired effect, it is often essential to scale the covariates, ensuring their variations are brought to a comparable level. The new grouping structure developed in this paper remains stable under such scaling operations. 

To harness the potential of multivariate FDA in analyzing homogeneity, we propose a novel grouped multiple functional regression framework. The ordinary multiple functional linear regression model (see also \cite{chiou2016multivariate} and \cite{mahzarnia2022multivariate}) is given below. For the $n$-th sample, $y_n\in\mathbb{R}$ represents the scalar response variable, and $X_{nj}(t)\in L^2[0,1]$ represents the $j$-th functional covariate, 
\begin{equation}
\label{ordinary_model}
y_n=\beta_0+\sum_{j=1}^p\int_0^1 X_{nj}(t)\beta_j(t)\,dt+\epsilon_n,\\\ \mbox{E}\epsilon_n=0,\ \mbox{Var}(\epsilon_n)=s^2.
\end{equation} 
This model faces two primary limitations. First, it does not account for the homogeneity of different covariates. Second, the estimation error can be significant when $p$ is large. Estimating a functional regression model typically involves dimension reduction by projecting both the functional coefficients and covariates onto a finite number of basis functions, using projection scores to estimate the truncated functional coefficients. If all functions are projected onto $D$ orthonormal basis functions, the ordinary model would require estimating $pD + 1$ unknown coefficients, leading to formidable high estimation error when $p$ is large. 

The new grouped multiple functional regression model developed here addresses these limitations. If covariates $X_{ni}(t)$ and $X_{nj}(t)$ belong to the same group, the coefficient functions $\beta_i(t),\beta_j(t)$ share the same ``shape" such that $\beta_i(t)=\mbox{const.}\beta_{j}(t)$, where ``const." represents a generic constant. This grouping criterion is clearly more inclusive than the equal coefficient fusion criterion in existing literature, because equal coefficient functions share the same shape but coefficients with common shape are not necessarily equal. In this grouping structure, there is a template coefficient function for each group, and the coefficient functions of all covariates in the same group are proportional to the associated template function. Given the grouping structure $\bm{\delta}=\{\delta_1,\ldots,\delta_K\}$ including $K$ groups, in which $\delta_k\cap\delta_{k'}=\emptyset$ for $k\ne k'$ and $\cup_{k=1}^K\delta_k=\{1,\ldots,p\}$, the new grouped multiple functional regression model is developed as follows
\begin{equation}
\label{simple_model}
y_n=\beta_0+\sum_{k=1}^{K}\sum_{j\in\delta_k}f_{j}\int_0^1 X_{nj}(t)\alpha_k(t)\,dt+\epsilon_n.
\end{equation}
In this model, the $p$ functional covariates are aggregated into $K$ groups. The within-group homogeneity is characterized by the template functions $\{\alpha_k(t)\colon k=1,\ldots,K\}$, and the scale coefficients $\{f_j\in\mathbb{R}\colon j=1,\ldots,p\}\in\mathbb{R}^p$ explain the discrepancy of coefficient magnitude. For this grouped model, we need to estimate $\beta_0$, $\{f_j\colon j=1,\ldots,p\}$ and $\{\alpha_k(t)\colon k=1,\ldots,K\}$.  After dimension reduction, the number of unknown coefficients is reduced to $KD + p + 1$, representing a reduction of $(p - K)D - p$ coefficients. The challenge lies in detecting the unknown grouping structure $\bm{\delta}=\{\delta_1,\ldots,\delta_K\}$. To address this, we develop a new regularization approach involving a new pairwise ``shape misalignment" penalty function. The new regularization approach shrinks small shape misalignment of coefficient functions, and the covariates with sufficiently small coefficient shape misalignment are then grouped together.  The grouping process is entirely data-driven and applicable to a wide range of cases. We also derive the conditions that ensure consistency of the detected grouping structure and investigate the asymptotic properties of the model estimates.

It is worth noting that,  when $K=1$, the new grouped model (\ref{simple_model}) can be viewed as an extension of the matrix-variate regression model $y_n=\bm{\alpha}^{\top}X_n\bm{\beta}+\epsilon_n$, where $X_n$ is some  $p\times D$ matrix-type covariate, $\bm{\alpha}$ is the $p\times 1$ row coefficients, and $\bm{\beta}$ is the $D\times 1$ column coefficients (see, e.g.,  \cite{li2010dimension}, \cite{hung2013matrix}, \cite{zhou2013tensor} and \cite{ding2018matrix}). To see this, note that $\bm{\beta}$ corresponds to $\alpha_1(t)$, $\bm{\alpha}$ corresponds to $\{f_j\colon j\ge1\}$, and each row of $X_n$ represents a finite-dimensional functional covariate, with the dimension determined by the number of columns in $X_n$. Clearly, compared to our grouped model, the limitation of this matrix-variate regression model is that all rows of $X_n$ share the same template coefficient $\bm{\beta}$, which can be overly restrictive in practice. On the contrary, the new grouped model allows for multiple groups and templates. We demonstrate that when not all covariates belong to the same group, employing matrix-variate regression can lead to a significant loss of predictive power due to model misspecification. This loss arises because the underlying assumptions of the matrix-variate regression model do not adequately capture the true relationships and variations among the covariates, resulting in inaccurate predictions and potentially misleading conclusions. Consequently, accurately grouping covariates becomes essential for enhancing the model’s performance and reliability. 
 
The rest of the paper is organized as follows. In Section \ref{s2}, we develop the new grouped multiple functional regression framework, including the shape-based grouping structure and the group detection procedure. Theoretical properties are developed in Section \ref{thr}. Finite-sample properties of the proposed methods are investigated through simulation studies in Section \ref{sim}. A case study on sugar quality evaluation is provided in Section \ref{rd}. Section \ref{con} concludes the paper. Technical proofs and additional simulation results are given in the Supplementary Material.

\section{Grouped Multivariate Functional Regression Based on Coefficient Shape Homogeneity}
\label{s2}
\subsection{Shape-based Grouping Structure}
Suppose that there are $p$ functional covariates $\{X_{j}(t)\colon j=1,\ldots,p\}$ and a scalar response $y$. For each sample $n\in\{1,2,\ldots,N\}$, we assume that $X_{nj}(t)\in L^2[0,1]$ and $y_n\in\mathbb{R}$. In the functional space $L^2[0,1]$, the inner product is defined as $\langle x,y \rangle=\int_0^1x(t)y(t)\,dt$, and the $L^2$-norm is defined as $\|x\|^2=\int_0^1x^2(t)\,dt<\infty$. The ordinary multivariate functional linear regression model without coefficient constraints is given by (\ref{ordinary_model}), where $\beta_0\in\mathbb{R}$ and $\beta_j(t)\in L^2[0,1]$ for $j=1,\ldots,p.$ Assume that the $p$ functional covariates are aggregated into $K$ disjoint groups $\bm{\delta}=\{\delta_1,\ldots,\delta_K\}$ such that $\delta_k\cap\delta_{k'}=\emptyset$ for $k\ne k'$ and $\cup_{k=1}^K\delta_k=\{1,\ldots,p\}$, then in our grouped model, the coefficient functions are defined in the restricted space $\bm{\Theta}_{\delta}\overset{\vartriangle}=\{(\beta_j(t)\colon j\ge1)\colon \beta_j(t)=f_j\alpha_{k}(t)\ \mbox{for}\ j\in\delta_k,\ f_j\in\mathbb{R},\ \alpha_k(t)\ne0,\ k=1,\ldots,K\}$, where $\{f_j\in\mathbb{R}\colon j=1,\ldots,p\}$ are termed scale coefficients and $\{\alpha_k(t)\in L^2[0,1]\colon k=1,\ldots,K\}$ are termed template coefficient functions. To distinguish different groups, we assume that $\langle\alpha_{k},\alpha_{k'}\rangle/\|\alpha_{k}\|\|\alpha_{k'}\|\ne1$ for any $k\ne k'$. In this paper, the grouping structure $\bm{\delta}$ is assumed unknown, and a new regularization approach is developed for group detection. This grouping structure has two main advantages. First it is more inclusive than equal coefficient grouping. That is because if $\beta_i(t)=\beta_j(t)$ then the two coefficients share the same template, however $\beta_i(t)$ and $\beta_j(t)$ are not necessarily equal even if they share the same template. Another advantage of this new grouping structure is its {\it invariance} to covariate scaling, because two coefficients in the same group will retain the same template even when their associated covariates are scaled differently. Specifically, for $i,j\in\delta_k$ and $c,c'\in\mathbb{R}\setminus\{0\}$, if the covariates $X_{i}(t)$ and $X_{j}(t)$ are scaled to $cX_{i}(t)$ and $c'X_{j}(t)$, the associated  coefficient functions will be scaled to $c^{-1}\beta_i(t)$ and ${c'}^{-1}\beta_{j}(t)$, while still sharing the same template.

Since this paper focuses on identifying homogeneous groups of covariates, we assume that  $f_j \neq 0$  for all $j$. Covariates associated with zero coefficients cannot explain the response, and including them only increases estimation error. To circumvent this difficulty, a pre-selection procedure based on the group-LASSO technique (see e.g., \cite{yuan2006model} and \cite{mahzarnia2022multivariate}) is implemented to remove insignificant functional covariates. The proposed grouping procedure is then applied to the remaining covariates.
\subsection{Group Detection by Coefficient Shape Alignment}
\label{thresh}
In this section, we develop a regularization approach for group detection, termed coefficient shape alignment. Suppose that there exist some orthonormal basis functions $\{\nu_d(t) \colon d \in\mathbb{N}\}$ such that the functional covariates and coefficients admit the following basis representations
\begin{align*}
\beta_j(t)=\sum_{d=1}^{\infty}b_{jd}\nu_d(t), \ \ X_{nj}(t)=\sum_{d=1}^\infty\xi_{nj,d}\nu_d(t).
\end{align*} 
With this representation, the functional model (\ref{ordinary_model}) can be rewritten as
\begin{align*}
y_n=\beta_0+\sum_{j=1}^p\sum_{d=1}^\infty \xi_{nj,d}b_{jd}+\epsilon_n.
\end{align*} 
Here we define the coefficient shape misalignment between covariate $i$ and $j$ as $M_{ij,dd'}\overset{\vartriangle}= b_{id}b_{jd'}-b_{jd}b_{id'},\ 1\le d<d'<\infty.$ Notationally, let $\bm{M}^\infty_{ij}=(\cdots,M_{ij,dd'},\cdots)$ be the array composed of all $M_{ij,dd'}$'s. If two covariates $X_{ni}(t)$ and $X_{nj}(t)$ are in the same group, their associated coefficient functions should be proportional, leading to $\bm{M}^\infty_{ij}=\bm{0}$, where $\bm{0}=(\ldots,0,\ldots)$. Therefore, we develop a pairwise shape misalignment penalty to regularize the ordinary model estimates for group detection. Specifically, we first minimize the following objective function, 
\begin{equation}
\label{loss}
S(\{b_{jd}\colon 1\le j\le p,\ d\ge1\},\lambda)=\frac{1}{2}\sum_{n=1}^N\left(y_n-\sum_{j=1}^p\sum_{d=1}^\infty\xi_{nj,d}b_{jd}\right)^2+\sum_{i<j}J_{\lambda}(\|\bm{M}^\infty_{ij}\|),
\end{equation}
where $J_{\lambda}(\cdot)$ is a non-convex penalty function, $\lambda$ is a tuning parameter, and $\|\cdot\|$ denotes the $L^2$ norm. 

However, it is impractical to minimize the objective function (\ref{loss}) due to its involvement of infinite-dimensional arguments. Therefore, we propose to minimize the following truncated objective function instead
\begin{equation}
\label{trunloss}
S_D(\bm{B},\lambda)=\frac{1}{2}\sum_{n=1}^N\left(y_n-\sum_{j=1}^p\sum_{d=1}^D\xi_{nj,d}b_{jd}\right)^2+\sum_{i<j}J_{\lambda}(\|\bm{M}_{ij}\|),
\end{equation}
where $\bm{B}=(b_{11},\ldots,b_{1D},\ldots,b_{p1},\ldots,b_{pD})^{\top}$, $\bm{M}_{ij}=(M_{ij,dd'}\colon 1\le d<d'\le D)$, and $D$ is selected such that the approximation error $\sum_{j=1}^p\sum_{d\ge D+1}\xi^2_{nj,d}$ is negligible. For example, we can combine all covariate functions and apply functional principal component analysis to the combined set, and select $D$ such that the overall cumulative proportion of variance explained exceeds a pre-specified threshold (e.g., 85\% -- 95\%). This selection process allows us to assume that the major components of functional covariates account for the significant variation in the response (see \cite{cai2006prediction}, where it is assumed that $b_{jd}\le\mbox{const.}\,d^{-\beta}$ for some $\beta>1$). Under these circumstances, the shape discrepancy can be effectively captured by $\{\bm{M}_{ij}\colon 1\le i<j\le p\}$ when the approximation error is small. 

The regularized estimate of $\bm{B}$ is defined as $\widehat{\bm{B}}(\lambda)=\arg\min_{\bm{B}}S_D(\bm{B},\lambda).$ 
We then compute $\{\widehat{\bm{M}}_{ij}(\lambda)\colon i<j\}$ from 
$\widehat{\bm{B}}(\lambda)$ as 
$$\widehat{M}_{ij,dd'}(\lambda)\overset{\vartriangle}= \hat{b}_{id}(\lambda)\hat{b}_{jd'}(\lambda)-\hat{b}_{jd}(\lambda)\hat{b}_{id'}(\lambda),$$ and specify a threshold $\tilde{\lambda}$ to truncate $\|\widehat{\bm{M}}_{ij}(\lambda)\|/(\|\widehat{\bm{B}}_i(\lambda)\|\|\widehat{\bm{B}}_j(\lambda)\|)$ to aggregate functional covariates, where $\widehat{\bm{B}}_j(\lambda)=(\hat{b}_{j1}(\lambda),\ldots,\hat{b}_{jD}(\lambda))^{\top}$. 
For example, $\beta_{i_1}(t),\beta_{i_2}(t),\beta_{i_3}(t)$ are grouped together when $\|\widehat{\bm{M}}_{ii'}(\lambda)\|\le\tilde{\lambda}\|\widehat{\bm{B}}_{i}(\lambda)\|\|\widehat{\bm{B}}_{i'}(\lambda)\|$ for each pair in $i_1,i_2,i_3$. 
We propose to threshold $\|\widehat{\bm{M}}_{ij}(\lambda)\|/(\|\widehat{\bm{B}}_i(\lambda)\| \|\widehat{\bm{B}}_j(\lambda)\|)$ instead of $\|\widehat{\bm{M}}_{ij}(\lambda)\|$, because $\|\widehat{\bm{M}}_{ij}(\lambda)\|$ may be small if either $\|\widehat{\bm{B}}_i(\lambda)\|$ or $\|\widehat{\bm{B}}_j(\lambda)\|$ is small. By scaling the misalignment before thresholding, we ensure that the common threshold $\tilde{\lambda}$ works effectively for all $\{\widehat{\bm{M}}_{ij}(\lambda)\colon i<j\}$. 

The function $J_\lambda(\cdot)$ is important for group detection. The LASSO-type penalty (see \cite{tibshirani1996regression}) applies the same thresholding to all $\|\bm{M}_{ij}\|$, and thus leads to biased estimates. Non-convex functions are usually used to solve this limitation of LASSO-type penalty. Three concave penalty functions are considered here: the truncated LASSO penalty $J_{\lambda}(x)=\min\{\lambda|x|,\gamma\lambda^2\}$ (see \cite{shen2010grouping}), the minimax concave penalty $J_{\lambda}(x)=\min\{\lambda|x|-{x^2}/{2\gamma},\gamma\lambda^2/2\}$ (MCP, see \cite{zhang2010nearly}), and the smoothly clipped absolute deviation (SCAD) penalty $J(x)=\int_0^{|x|}\lambda\min\{1,\frac{(\gamma\lambda-t)_+}{(\gamma-1)\lambda}\}dt$ (see \cite{fan2001variable}). Note that the misalignment term is quadratic, which complicates the optimization problem. To solve this issue, we employ the linearized ADMM algorithm, which we discuss below.
\begin{remark}
The tuning parameters $\lambda$ and $\tilde{\lambda}$ can be selected using cross validation techniques. Specifically, first choose some candidates of $\lambda, \tilde{\lambda}$ to detect a set of candidate grouped models, and then apply the cross validation procedure to estimate the prediction mean squared errors of the candidate grouped models. The tuning parameters associated with the model giving the best prediction are selected.
\end{remark}
\begin{remark}
Different covariates exhibit display different variation patterns and it would be more efficient to employ different sets of basis functions to represent different functions. However, if different sets of basis are employed, two coefficient functions are not aligned when the associated scores are aligned. Since a key objective is to detect the grouping structure based on coefficient shape homogeneity, it is essential to use the same set of basis functions for all covariates.
\end{remark}

\subsection{Computation with Linearized ADMM Algorithm}
\label{s3}
The alternating direction method of multipliers (ADMM) algorithm (see \cite{glowinski1975approximation} and \cite{gabay1976dual}) is typically employed to solve optimization problems with linear equality constraints. However, the challenge in this context is that the penalty considered involves the quadratic terms $\{\bm{M}_{ij}\colon i < j\}$, making it infeasible to develop an equivalent formulation with linear constraints for the objective function  (\ref{trunloss}). To solve this issue, we employ the linearized ADMM algorithm. 

Since the intercept is not relevant to the group detection process, we omit it for simplicity. Let
$$H_1(\bm{B})=\frac{1}{2}\sum_{n=1}^N\left(y_n-\sum_{j=1}^p\bm{\xi}^{\top}_{nj}\bm{B}_{j}\right)^2,\qquad H_2(\bm{M})=\sum_{i<j}J_{\lambda}(\|\bm{M}_{ij}\|),$$
where $\bm{\xi}_{nj}=(\xi_{nj,1},\ldots,\xi_{nj,D})^{\top}$, $\bm{B}_j=(b_{j1},\ldots,b_{jD})^{\top}$, and $\bm{M}=\{\bm{M}_{ij}\colon 1\le i<j\le p\}$, 
then the unconstrained optimization problem (\ref{trunloss}) is equivalent to the following constrained optimization problem
\begin{align}
\label{loss2}
&\min_{\bm{B},\bm{M}}H(\bm{B},\bm{M})=\min_{\bm{B},\bm{M}} H_1(\bm{B})+H_2(\bm{M}),\nonumber\\
&\hspace{1cm} \mbox{subject to}\ M_{ij,dd'}=b_{id}b_{jd'}-b_{jd}b_{id'},\ 1\le i<j\le p,\ 1\le d< d'\le D.
\end{align}
Notationally, let $F(\bm{B},\bm{M})=\bm{0}$ comprises all the constraints in (\ref{loss2}). By the ADMM algorithm, the regularized estimates of $\bm{B},\bm{M}$ can be obtained by minimizing
\begin{align*}
L_{\theta}(\bm{B},\bm{M},\bm{u})&=H(\bm{B},\bm{M})+\sum_{i<j}\sum_{d<d'}u_{ij,dd'}((b_{id}b_{jd'}-b_{jd}b_{id'})-M_{ij,dd'})\\
&\hspace{2cm}+\frac{\theta}{2}\sum_{i<j}\sum_{d<d'}((b_{id}b_{jd'}-b_{jd}b_{id'})-M_{ij,dd'})^2
\end{align*}
over $\bm{B}$ and $\bm{M}$. Given $\bm{B}$ and multipliers $\bm{u}=\{u_{ij,dd'},i<j,d<d'\}$, we update $\bm{M}$ by minimizing $L_{\theta}(\bm{B},\bm{M},\bm{u})$ over $\bm{M}$. The optimization problem is equivalent to minimizing the following function over $\bm{M}$,
\begin{equation*}
\frac{\theta}{2}\sum_{i<j}\sum_{d<d'}(M_{ij,dd'}-(b_{id}b_{jd'}-b_{jd}b_{id'})-\theta^{-1}u_{ij,dd'})^2+\sum_{i<j}J_{\lambda}(\|\bm{M}_{ij}\|).
\end{equation*}
Notionally, let $a_{ij,dd'}=(b_{id}b_{jd'}-b_{jd}b_{id'})+\theta^{-1}u_{ij,dd'}$. For the truncated LASSO penalty, the solution of $\bm{M}_{ij}$ is
\begin{equation}
\label{l1}
\bm{M}_{ij}=\left\{
\begin{array}{ccc}
\bm{a}_{ij}\left(1-\frac{\lambda}{\theta
\|\bm{a}_{ij}\|}\right)_+,& &\mbox{if }\|\bm{a}_{ij}\|\le\lambda\left(\gamma+\frac
{1}{2\theta}\right).\\
\bm{a}_{ij},& &\mbox{if }\|\bm{a}_{ij}\|>\lambda\left(\gamma+\frac{1}{2\theta}\right).
\end{array}
\right.
\end{equation}
For the MCP, the solution is
\begin{equation}
\label{MCP}
\bm{M}_{ij}=\left\{
\begin{array}{ccc}
\frac{\bm{a}_{ij}\left(1-\frac{\lambda}{\theta\|\bm{a}_{ij}\|}\right)_+}{1-\frac{1}{\gamma\theta}},& &\mbox{if }\|\bm{a}_{ij}\|\le\gamma\lambda.\\
\bm{a}_{ij},& &\mbox{if }\|\bm{a}_{ij}\|>\gamma\lambda.
\end{array}
\right.
\end{equation}
For the SCAD penalty, the solution is
\begin{equation}
\label{SCAD}
\bm{M}_{ij}=\left\{
\begin{array}{ccc}
\bm{a}_{ij}\left(1-\frac{\lambda}{\theta\|\bm{a}_{ij}\|}\right)_+,& &\mbox{if }\|\bm{a}_{ij}\|\le\lambda\left(1+\frac{1}{\theta}\right).\\
\frac{\bm{a}_{ij}\left(1-\frac{\gamma\lambda}{\theta(\gamma-1)\|\bm{a}_{ij}\|}\right)_+}{1-\frac{1}{\theta(\gamma-1)}},& &\mbox{if }\lambda\left(1+\frac{1}{\theta}\right)<\|\bm{a}_{ij}\|\le\gamma\lambda.\\
\bm{a}_{ij},& &\mbox{if }\|\bm{a}_{ij}\|>\gamma\lambda.
\end{array}
\right.
\end{equation}
The sub-problem of minimization over $\bm{B}$ given $\bm{M}$ and $\bm{u}$ is complicated due to the non-linearity of the constraint functions $F(\bm{B},\bm{M})$. To avoid this intractable non-linear sub-problem, we employ linearization techniques (see, e.g.,  
\cite{benning2015preconditioned} and \cite{latorre2019fast}). At iteration $m+1$, we replace the constraint function $F(\bm{M},\bm{B})$ with its first-order Taylor expansion around the value of $\bm{B}$ at the previous iteration $m$, denoted by $\bm{B}^{(m)}$,  
$F(\bm{M},\bm{B})\approx F(\bm{M},\bm{B}^{(m)})+\partial_{\bm{B}}F(\bm{M},\bm{B}^{(m)})(\bm{B}-\bm{B}^{(m)})\overset{\vartriangle}= \widetilde{F}(\bm{M},\bm{B}).$
For $1\le j\le p$, define
$\bm{\Xi}_j=\left(\bm{\xi}_{1j},\ldots,\bm{\xi}_{Nj}\right)^{\top}$, $\bm{\Xi}=(\bm{\Xi}_1|\ldots|\bm{\Xi}_p)$. The augmented Lagrangian function is replaced by the following approximation
\begin{align*}
\widetilde{L}_{D,\theta}(\bm{B},\bm{M},\bm{u})
&=\frac{1}{2}\|\bm{y}-\bm{\Xi}\bm{B}\|^2+\bm{u}^{\top}\widetilde{F}(\bm{M},\bm{B})+\frac{\theta}{2}\|\widetilde{F}(\bm{M},\bm{B})\|^2_2+c(\bm{M}),
\end{align*}
where $c(\bm{M})$ is related to $\bm{M}$ only. The minimizer of $\bm{B}$ given $\bm{M}$ and $\bm{u}$ of the above approximated objective function is
\begin{align}
\bm{B}&=(\bm{\Xi}^{\top}\bm{\Xi}+\theta\partial_{\bm{B}}F(\bm{M},\bm{B}^{(m)})^{\top}\partial_{\bm{B}}F(\bm{M},\bm{B}^{(m)}))^{-1}
\{\bm{\Xi}^{\top}\bm{y}-\theta\partial_{\bm{B}}F(\bm{M},\bm{B}^{(m)})^{\top}F(\bm{M},\bm{B}^{(m)})\nonumber\\
\label{SB}
&\hspace{1cm}-\partial_{\bm{B}}F(\bm{M},\bm{B}^{(m)})^{\top}\bm{u}+\theta\partial_{\bm{B}}F(\bm{M},\bm{B}^{(m)})^{\top}\partial_{\bm{B}}F(\bm{M},\bm{B}^{(m)})\bm{B}^{(m)}\}.
\end{align}
Based on the above discussion, we summarize the algorithm as follows:
\begin{algorithm}[ht]
 \caption{Linearized ADMM algorithm}
  \begin{algorithmic}[1]
   \State {Initialize estimates $\bm{B}^{(0)}$ (e.g., {ordinary least squares estimate of $\bm{B}$}) and set $\bm{u}^{(0)}=\bm{0}$.}
  \While{{convergence criterion is not met}}
  \State  {Given $\bm{B}^{(m)}$ and $\bm{u}^{(m)}$, calculate $\bm{M}^{(m+1)}$ using equations (\ref{l1}), (\ref{MCP}), and (\ref{SCAD}).}
   \State  {Given $\bm{M}^{(m+1)}$ and $\bm{u}^{(m)}$, calculate $\bm{B}^{(m+1)}$ using equation (\ref{SB}).}
   \State {Update the multipliers $u^{(m+1)}_{ij,dd'}=u^{(m)}_{ij,dd'}+\theta(b^{(m+1)}_{id}b^{(m+1)}_{jd'}-b^{(m+1)}_{jd}b^{(m+1)}_{id'}-M^{(m+1)}_{ij,dd'})$.}
  \EndWhile
      \State \Return {$\bm{B}^{(m+1)}$.}
  \end{algorithmic}
  \label{al1}
\end{algorithm}

\subsection{Grouped Model Estimation}
\label{est}
After detecting the unknown grouping structure, the next step is to establish and estimate the grouped model. Given the detected grouping structure $\hat{\bm{\delta}}=(\hat{\delta}_1,\ldots,\hat{\delta}_{K})$, the grouped model is 
$$y_n=\beta_0+\sum_{k=1}^{K}\sum_{j\in\hat{\delta}_k}f_{j}\langle X_{nj},\alpha_k\rangle+\epsilon_n,$$ 
where $\beta_{j}(t)=f_j\alpha_k(t)$ if $j\in\hat{\delta}_k$. 
Define $g_k(i)$ as the index of the $i$-th covariate in the $k$-th group, and $|\hat{\delta}_k|$ as the cardinality (number of covariates) of group $\hat{\delta}_k$, and let $\bm{z}_{nk}=(\bm{\xi}_{ng_k(1)},\ldots,\bm{\xi}_{ng_k({|\hat{\delta}_k|)}})^{\top}$, $\bm{f}_k=(f_{g_k(1)},\ldots,f_{g_k({|\hat{\delta}_k|})})^{\top}$ and $\bm{\alpha}_k=(a_{k1},\ldots,a_{kD})^{\top}=(\langle\alpha_k,\nu_1\rangle,\ldots,\langle\alpha_k,\nu_D\rangle)^{\top}$ denoting the template coefficient scores of group $k$. The grouped model is rewritten as
\begin{equation}
\label{trunmodel}
y_n=\beta_0+\sum_{k=1}^K\bm{f}^{\top}_k\bm{z}_{nk}\bm{\alpha}_k+\epsilon_n+e_n^D,
\end{equation}
where  
$e_n^D=\sum_{k=1}^K\sum_{j\in\hat{\delta}_k}\sum_{d\ge D+1}f_{j}a_{kd}\xi_{nj,d}$ is the truncation error. Although $\{\bm{\alpha}_k,\bm{f}_k\colon k\ge1\}$ are not identifiable (since $\{\bm{\alpha}_k,\bm{f}_k\colon k\ge1\}$ and $\{c_k\bm{\alpha}_k,\bm{f}_k/c_k\colon k\ge1\}$ lead to the same model for arbitrary non-zero constants $\{c_k\colon k\ge1\}$), the products $\{\bm{f}_k\otimes\bm{\alpha}_k\colon k\ge1\}$ are identifiable. Here ``$\otimes$" denotes the Kronecker product. Since $f_{g_k(i)}\bm{\alpha}_k=\bm{B}_{g_k(i)}$, we only need to estimate $\{\bm{\alpha}_k,\bm{f}_k\colon k\ge1\}$.

Using the least squares method, the estimates are obtained by solving
\begin{equation*}
\{\hat{\beta}_0,\widehat{\bm{\alpha}}_k,\widehat{\bm{f}}_k\colon k\ge1\}=\arg\min_{\beta_0,\{\bm{\alpha}_k,\bm{f}_k\colon k\ge1\}}\frac{1}{2}\sum_{n\ge1}\left(y_n-\beta_0-\sum_{k\ge1}\bm{f}^{\top}_k\bm{z}_{nk}\bm{\alpha}_k\right)^2.
\end{equation*}
Although the above objective function is not jointly linear in $\{\bm{\alpha}_k,\bm{f}_k\colon k\ge1\}$, it is linear in $\{\bm{f}_k\colon k\ge1\}$ or $\{\bm{\alpha}_k\colon k\ge1\}$ individually. This motivates an iterative approach, updating $\{\bm{f}_k\colon k\ge1\}$ and $\{\bm{\alpha}_k\colon k\ge1\}$ alternately. In each iteration, either $\{\bm{f}_k\colon k\ge1\}$ or $\{\bm{\alpha}_k\colon k\ge1\}$ is updated while keeping the other one fixed. This iterative algorithm is known as block relaxation algorithm (see e.g., \cite{lazar2010numerical}). Notationally, define 

\begin{align*}
\bm{Z}_{n}=\left[\begin{array}{cccc}
\bm{z}_{n1}&&&\\
&\bm{z}_{n2}&&\\
&&\ddots&\\
&&&\bm{z}_{nK}
\end{array}
\right], \ \
\bm{F}=\left[\begin{array}{c}
\bm{f}_1\\
\bm{f}_2\\
\vdots\\
\bm{f}_{K}\\
\end{array}
\right], \ \
\bm{A}=\left[\begin{array}{c}
\bm{\alpha}_1\\
\bm{\alpha}_2\\
\vdots\\
\bm{\alpha}_{K}\\
\end{array}
\right],
\end{align*}
and the truncated grouped model (\ref{trunmodel}) is then rewritten as $y_n=\beta_0+\bm{F}^{\top}\bm{Z}_n\bm{A}+\epsilon_n+e_n^D$. 
The iterative estimation procedure is summarized as follows: 
\begin{algorithm}[ht]
 \caption{Iterative estimation algorithm}
  \begin{algorithmic}[1]
   \State {Initialise $\bm{F}^{(0)},{\bm{A}}^{(0)}$ and $\beta^{(0)}_0$.}
 \While{{convergence criterion is not met}}
  \State  {Fix $\beta_0^{(m)}$, ${\bm{F}}^{(m)}$ and update ${\bm{A}}^{(m+1)}=\arg\min_{\bm{A}}\sum_{n\ge1}(y_n-\beta_0^{(m)}-{\bm{F}^{(m)}}^{\top}\bm{Z}_n\bm{A})^2$.}
   \State  {Fix $\beta_0^{(m)}$, ${\bm{A}}^{(m+1)}$ and update ${\bm{F}}^{(m+1)}=\arg\min_{\bm{F}}\sum_{n\ge1}(y_n-\beta_0^{(m)}-\bm{F}^{\top}\bm{Z}_n\bm{A}^{(m+1)})^2$.}
     \State  {Fix ${\bm{F}}^{(m+1)}$, ${\bm{A}}^{(m+1)}$ and update 
      $\beta_0^{(m+1)}=\arg\min_{\beta_0}\sum_{n\ge1}(y_n-\beta_0-\bm{F}^{(m+1)^{\top}}\bm{Z}_n\bm{A}^{(m+1)})^2.$}
   \EndWhile
 \State \Return $\bm{A}^{(m+1)}, \bm{F}^{(m+1)}, \beta_0^{(m+1)}$.
  \end{algorithmic}
  \label{al1}
\end{algorithm}
\section{Theoretical Properties}
\label{thr}
\subsection{Consistency of Group Detection}
\label{consistency}
In this section, we investigate the consistency properties of the detected grouping structure. We establish the necessary conditions on the tuning parameters, shape misalignment, and the dimension $D$ required to accurately recover the true grouping structure. We demonstrate that, under these regularity conditions, there exists a local minimizer of the objective function (\ref{trunloss}) around the true coefficients such that the associated grouping structure coincides with the true grouping structure asymptotically almost surely. 
Before presenting the main theoretical results, we first introduce the following assumptions:
\begin{itemize}
\item[(A1)] For arbitrary $n,j$, $|\xi_{nj,d}|\le U_jd^{-r_j}$ with  $r_j>1/2$.
\item[(A2)] For arbitrary $j$, $|b_{jd}|\le\mbox{const.}d^{-r_{\beta_j}}$ with $r_{\beta_j}>1/2$.
\item[(A3)] $J_\lambda(t)$ is a non-decreasing and concave function for $t\in [0, \infty)$ and $J_\lambda(0) = 0$. There exists a constant $\gamma>0$ such that $J_\lambda(t)$ is constant for all $t\ge \gamma\lambda\ge0$. The gradient $J'_\lambda(t)$ exists and is continuous except for a finite number of $t$ and $\lim_{t\to0+}J'_\lambda(t) = \lambda$.
\item[(A4)] $\bm{\epsilon}=(\epsilon_1,\ldots,\epsilon_n)$ follows a sub-Gaussian distribution, meaning there exists $C_1>0$ so that $P(|\bm{s}^{\top}\bm{\epsilon}|>\|\bm{s}\|x)\le2\exp(-C_1x^2)$ for any vector $\bm{s}\in\mathbb{R}^n$ and $x>0$.
\end{itemize}
In our setting, $p$ is fixed and $D$ is allowed to increase to infinity.
Assumption (A1) and (A2) hold for functions in $L^2[0,1]$ (see also \cite{hall2007methodology}, \cite{hall2006properties} and \cite{jiao2023functional}). 
Assumption (A3) holds for all the three penalties considered in this paper. Assumption (A4) provides the theoretical foundation for bounding the probability of incorrect grouping.

Let
$\{\bm{M}^0_{ij}\colon i<j\}$ represent the coefficient shape misalignment of the underlying true coefficient scores $\bm{B}^0$, and define $\tau_N=\sqrt{N^{-1}\log N}$. We develop the following theorem about the consistency of the detected grouping structure.
\begin{theorem}
\label{thm2}
Suppose that Assumptions (A1) --- (A4) hold. If for arbitrary $k\ne k'$, and $i\in\delta_k$, $j\in\delta_{k'}$, it satisfies that 
\begin{equation*}
\label{con1}
\|\bm{M}^0_{ij}\|-2\{\tau_N^2+\tau_N(\|\bm{B}^0_i\|+\|\bm{B}^0_j\|)\}>\max\{\gamma\lambda,\tilde{\lambda}(\|\bm{B}^0_i\|+\tau_N)(\|\bm{B}^0_j\|+\tau_N)\},
\end{equation*}
and additionally if
\begin{equation*}
\label{con2}
\lambda\tilde{\lambda}\sum_{k\ge1}\sum_{\{i,j\}\in\delta_k}\|\bm{B}^0_i\|\|\bm{B}^0_j\|(\log N)^{-1}\to\infty,
\end{equation*}
then there exists a local minimizer of (\ref{trunloss}) around $\bm{B}^0$ satisfying that $P(\hat{\bm{\delta}}\ne\bm{\delta})\le2pDN^{-1}$, where $\hat{\bm{\delta}}$ is the grouping structure associated with the minimizer.
\end{theorem}
Theorem \ref{thm2} establishes the consistency of the detected grouping structure, obtained through thresholding as discussed in Section \ref{thresh}.
Note that when $DN^{-1}\to0$, $P(\hat{\bm{\delta}}_m=\bm{\delta})\to1$, which indicates that there exists a local minimizer of (\ref{trunloss}) of which the associated grouping structure coincides with the true grouping structure asymptotically almost surely. The conditions not only involve the tuning parameters, but also the magnitude of the true coefficients. If $\|\bm{B}_i^0\|$ and $\|\bm{B}_j^0\|$ are overly large, the condition $\|\bm{M}^0_{ij}\|-2\{\tau_N^2+\tau_N(\|\bm{B}^0_i\|+\|\bm{B}^0_j\|)\}>\tilde{\lambda}(\|\bm{B}^0_i\|+\tau_N)(\|\bm{B}^0_j\|+\tau_N)$ may not hold. In other words, the shape discrepancy should be sufficiently pronounced relative to the coefficient magnitude for it to be detectable. In addition, as $\gamma$ increases, the influence of the concavity of $J_\lambda(\cdot)$ diminishes, potentially invalidating the condition $\|\bm{M}^0_{ij}\|-2\{\tau_N^2+\tau_N(\|\bm{B}^0_i\|+\|\bm{B}^0_j\|)\}>\gamma\lambda$. The concavity helps differentiate large shape misalignments from small ones, preventing over-shrinkage and ensuring that the true grouping structure is detected.

It is worth noting that $\tau_N$ is an important component of the non-asymptotic upper bound of the estimation error of the oracle estimator $\widehat{\bm{B}}^{or}=\arg\min_{\bm{B}\in\bm{\Theta}^D_\delta}\left\|\bm{y}-\beta_0\bm{1}_N-\bm{\Xi}\bm{B}\right\|^2,$ where $\bm{\Theta}^D_{\delta}\overset{\vartriangle}=\{(\bm{B}_j\colon j\ge1)\colon \bm{B}_j=f_j\bm{\alpha}_{k}\ \mbox{for}\ j\in\delta_k,\ f_j\in\mathbb{R},\ \bm{\alpha}_k\ne0,\ k=1,\ldots,K\}$ and $\bm{1}_N=(1,1,\ldots,1)^{\top}$. Denote the largest and smallest eigenvalues of the matrix $N^{-1}\bm{\Xi}^{\top}\bm{\Xi}$ by $\sigma_{max},\sigma_{min}$. We introduce the following result.
\begin{theorem}
\label{thm0}
Suppose that Assumptions (A1), (A2) and (A4) hold. Then with probability greater than $1-2pDN^{-1}$, we have 
$\|\widehat{\bm{B}}^{or}-\bm{B}^0\|\le{4\sqrt{pC_D}\sigma_{max}}{\sigma^{-2}_{min}}\tau_N,$
where $C_D=C^{-1}_1[\min_i\{U_i^{-2}t^{-1}_{2r_i}(D)\}]^{-1}$ and $t_{\alpha}(D)=\sum_{d=1}^Dd^{-\alpha}$.
\end{theorem}
\subsection{Asymptotic Properties of Model Estimates}
Let $X(t)\in L^p_H$ indicate that, for some $p>0$, a $H$-valued random function $X(t)$ satisfies $\mbox{E}\{\|X(t)\|^p\}<\infty$.
In this section, we assume that $X_j(t)\in L^2_H$ for each $j$ where $H=L^2[0,1]$. The samples $\{(y_n,X_{nj}(t)\colon j=1,\ldots,p),\ n\ge1\}$ are assumed to be $i.i.d.$ across $n$, and the covariates are independent of the random errors $\{\epsilon_n\colon n\ge1\}$. It is additionally assumed that $\{X_{nj}(t)\colon j\ge1,n\ge1\}$ and $\{y_n\in n\ge1\}$ are of mean zero without loss of generality, and the true grouping structure is detected. 

Denote $\{\hat{\bm{f}}_k,\hat{\bm{\alpha}}_k\colon k\ge 1\}$ to be the least squares estimates of $\{\bm{f}_k,\bm{\alpha}_k\colon k\ge 1\}$. As discussed, $\{\bm{f}_k,\bm{\alpha}_k\colon k\ge 1\}$ in model (\ref{trunmodel}) are not identifiable. To achieve identifiability, we normalize the estimates as $\hat{\bm{f}}^*_k=\hat{c}_k\hat{\bm{f}}_k,\ \hat{\bm{\alpha}}^*_k=\hat{\bm{\alpha}}_k/\hat{c}_k,$ where $\hat{c}_k=\mbox{sign}(\hat{a}_{k1})\|\hat{\bm{\alpha}}_k\|$ (see also \cite{ding2018matrix}). This ensures that the normalized template coefficient scores have unit norm. Notationally, define $\hat{\bm{\theta}}=(\hat{\beta}_0,\hat{\bm{f}}^{\top}_1,\ldots,\hat{\bm{f}}^{\top}_K,\hat{\bm{\alpha}}^{\top}_1,\ldots,\hat{\bm{\alpha}}^{\top}_K)^{\top}$, and there exists $\bm{\theta}_0=(\beta_0,\bm{f}^{\top}_{0,1},\ldots,\bm{f}^{\top}_{0,K},\bm{\alpha}^{\top}_{0,1},\ldots,\bm{\alpha}^{\top}_{0,K})^{\top}$ such that $\hat{\bm{\theta}}$ is consistent with $\bm{\theta}_0$, and define $\hat{\bm{\theta}}^*=(\hat{\beta}_0,\hat{\bm{f}}^{*{\top}}_1,\ldots,\hat{\bm{f}}^{*{\top}}_K,\hat{\bm{\alpha}}^{*{\top}}_1,\ldots,\hat{\bm{\alpha}}^{*{\top}}_K)^{\top}$ and $\bm{\theta}_0^*=(\beta_0,\bm{f}^{*{\top}}_{0,1},\ldots,\bm{f}^*_{0,K},\bm{\alpha}^{*{\top}}_{0,1},\ldots,\bm{\alpha}^{*{\top}}_{0,1})^{\top}$ as the normalized counterpart of $\hat{\bm{\theta}}$ and $\bm{\theta}_0$. 
Further, let $\bm{A}_0=(\bm{\alpha}^{\top}_{0,1},\ldots,\bm{\alpha}^{\top}_{0,K})^{\top}$, $\bm{F}_0=(\bm{f}^{\top}_{0,1},\ldots,\bm{f}^{\top}_{0,K})^{\top}$, $\bm{A}^*_0=(\bm{\alpha}^{*{\top}}_{0,1},\ldots,\bm{\alpha}^{*{\top}}_{0,K})^{\top}$, $\bm{F}^*_0=(\bm{f}^{*{\top}}_{0,1},\ldots,\bm{f}^{*{\top}}_{0,K})^{\top}$, $Q_n=(1\ \bm{A}_0^{\top}\bm{Z}^{\top}_n\ \ \bm{F}_0^{\top}\bm{Z}_n)$, which is a ${1\times (g+KD+1)}$ vector, $\bm{Q}^{\top}=(Q^{\top}_1\cdots Q^{\top}_N)$, $\Gamma_Q=\mbox{E}(Q^{\top}_nQ_n)$, and denote the gradient ${\partial\bm{\theta}^*}/{{\partial\bm{\theta}}^{\top}}$ at $\bm{\theta}_0$ as $\mathcal{G}_{\bm{\theta}_0}$, where
\begin{equation*}
\mathcal{G}_{\bm{\theta}_0}\overset{\vartriangle}=\left(
\begin{array}{cccccc}
1&\bm{0}&\bm{0}&\cdots&\bm{0}&\bm{0}\\
\bm{0}&c_1I_{|\delta_1|}&c^{-1}_1\bm{f}^*_{0,1}{\bm{\alpha}_{0,1}^*}^{\top}&\cdots&\bm{0}&\bm{0}\\
\bm{0}&\bm{0}&c^{-1}_1(I_D-\bm{\alpha}^*_{0,1}{\bm{\alpha}_{0,1}^*}^{\top})&\cdots&\bm{0}&\bm{0}\\
\vdots&\vdots&\vdots&\ddots&\vdots&\vdots\\
\bm{0}&\bm{0}&\bm{0}&\cdots&c_KI_{|\delta_K|}&c^{-1}_K\bm{f}^*_{0,K}{\bm{\alpha}_{0,K}^*}^{\top}\\
\bm{0}&\bm{0}&\bm{0}&\cdots&\bm{0}&c^{-1}_K(I_D-\bm{\alpha}_{0,K}^*{\bm{\alpha}_{0,K}^*}^{\top})\\
\end{array}
\right)
\end{equation*}
and $c_k=\mbox{sign}(a_{0,k1})\|\bm{\alpha}_{0,k}\|$ with $\bm{\alpha}_{0,k}=(a_{0,k1},\ldots,a_{0,kD})^{\top}$. We introduce the following assumptions.
\begin{itemize}
\item[(A5)] Define $\sigma_d(\cdot)$ as the $d$-th eigenvalue of some generic square matrix, where $\sigma_1(\cdot)\ge\sigma_2(\cdot)\ge\cdots$. There exists constants $\alpha_g>0$, $\alpha_m>1$ such that $D=o(N^{1/(2\alpha_m+\max\{\alpha_g,2\alpha_m\})})$, and for some $C_u>C_l>0$, the following holds
$$C_ld^{-\alpha_m}\le\sigma_d\left(\mbox{E}(\bm{Z}_n^{\top}\bm{F}_0\bm{F}_0^{\top}\bm{Z}_n)\right)\le\sigma_d\left(\Gamma_Q\right)\le C_ud^{-\alpha_m},$$
$$C_ld^{-\alpha_g}\le \sigma_d(\mathcal{G}^{\top}_{\bm{\theta}_0}\mathcal{G}_{\bm{\theta}_0})\le C_ud^{-\alpha_g}.$$
\item[(A6)] Let $g_{d_1d_2}$ denote the component of $\Gamma^{-1}_Q$ in the $d_1$-th row and the $d_2$-th column, and let $q_{nd}$ represent the $d$-th component of $Q_n$. It holds that 
$$\sum_{d_1,d_2,d_3,d_4=1}^{KD+p+1}\mbox{E}\left(q_{nd_1}q_{nd_2}q_{nd_3}q_{nd_4}\right)g_{d_1d_2}g_{d_3d_4}=o(ND^{-2}),$$
$$\sum_{d_1,\ldots,d_8=1}^{KD+p+1}\mbox{E}\left(q_{nd_1}q_{nd_3}q_{nd_5}q_{nd_7}\right)\mbox{E}\left(q_{nd_2}q_{nd_4}q_{nd_6}q_{nd_8}\right)g_{d_1d_2}g_{d_3d_4}g_{d_5d_6}g_{d_7d_8}=o(N^2D^2).$$
\end{itemize}
For elements in $L^2_H$, the sum of eigenvalues of its covariance operator should be finite, which necessitates $\alpha_m>1$ in Assumption (A5). Note that $\mbox{E}(\bm{Z}_n^{\top}\bm{F}_0\bm{F}_0^{\top}\bm{Z}_n)$ is a principal submatrix of $\mbox{E}\{Q_n^{\top}Q_n\}$, thus by the eigenvalue interlacing theorem, $\sigma_d\left(\mbox{E}(\bm{Z}_n^{\top}\bm{F}_0\bm{F}_0^{\top}\bm{Z}_n)\right)\le\sigma_d\left(\mbox{E}\{Q_n^{\top}Q_n\}\right)$. Here, we apply the martingale method (see e.g., \cite{hall2014martingale}) to develop the central asymptotic distribution of $\hat{\bm{\theta}}^*$. Note that this method is also employed in \cite{ghorai1980asymptotic} and \cite{muller2005generalized} for orthogonal series density estimates and generalized functional linear model estimates. Assumption (A6) is similarly applied in \cite{muller2005generalized}. The following theorem states the central asymptotic distribution of the normalized estimates $\hat{\bm{\theta}}^*$.
\begin{theorem}
\label{thm3}
Under Assumption (A2), (A5) and (A6), it holds that
\begin{equation*}
\frac{N(\hat{\bm{\theta}}^*-\bm{\theta}_0^*)^{\top}\Gamma(\hat{\bm{\theta}}^*-\bm{\theta}_0^*)-(KD+p+1)}{\sqrt{2(KD+p+1)}}\overset{d}\to\mathcal{N}(0,1),\qquad \mbox{as} \ N\to\infty,
\end{equation*}
where $\Gamma=\tilde{\sigma}^{-2}(\mathcal{G}_{\bm{\theta}_0}^{-1})^{\top}\Gamma_Q\mathcal{G}_{\bm{\theta}_0}^{-1}$, and
$$\tilde{\sigma}^{2}=s^2+\mbox{var}\left(\sum_{j=1}^p\sum_{d\ge D+1}\xi_{nj,d}b_{nj}\right)+\mbox{cov}\left(\sum_{j=1}^p\sum_{d\ge D+1}\xi_{nj,d}b_{nj},\sum_{j=1}^p\sum_{d=1}^D\xi_{nj,d}b_{nj}\right).$$
\end{theorem}
Theorem 3 gives the joint central asymptotic distribution of all the estimates. When the intercept $\beta_0$ is not of interest, $KD+p+1$ is reduced to  $KD+p$, which leads to the removal of the first row and column from $\mathcal{G}$ and the first column from $\bm{Q}$. Since $p$ is a fixed value and $D\to\infty$, Theorem 3 also holds when $KD+p+1$ is replaced with $KD$. 

For the scale and template coefficients, the central asymptotic distributions are also developed. Define 
\begin{align*}
&\bm{X}_{n}^a=\left(\int_0^1 X_{ng_1(1)}(t)\alpha_{0,1}(t)dt,\ldots,\int_0^1 X_{ng_1(|\delta_1|)}(t)\alpha_{0,1}(t)dt,\ldots,\right.\\
&\hspace{3cm}\left.\int_0^1 X_{ng_K(1)}(t)\alpha_{0,K}(t)dt,\ldots,\int_0^1 X_{ng_K(|\delta_K|)}(t)\alpha_{0,K}(t)dt\right)^{\top},
\end{align*}
$$\mathcal{G}_1=\mbox{diag}(c_1I_{|\delta_1|},\ldots,c_KI_{|\delta_K|}),\ {\Gamma}_a=\mbox{E}(\bm{X}_n^a\bm{X}_n^{aT}),$$
where 
$\alpha_{0,k}(t)=\sum_{d=1}^{\infty}a_{0,kd}\nu_d(t)$.  
Then we develop the following central asymptotic distribution for the estimates of normalized scale coefficients $\widehat{\bm{F}}_0$.
\begin{theorem}
\label{thm4}
Under Assumption (A2) and (A5), it holds that
$$\sqrt{N}(\widehat{\bm{F}}^*-\bm{F}_0^*)\overset{d}\to\mathcal{N}(0,s^2\mathcal{G}_1\Gamma^{-1}_a\mathcal{G}_1),\qquad \mbox{as} \ N\to\infty.$$
\end{theorem}
To establish the central asymptotic distribution for the estimates of the normalized template coefficient scores $\widehat{\bm{A}}^*$, first define 
$$\mathcal{G}_2=\mbox{diag}(c^{-1}_1(I_D-\bm{\alpha}^*_{0,1}{\bm{\alpha}_{0,1}^*}^{\top}),\ldots,c^{-1}_K(I_D-\bm{\alpha}^*_{0,K}{\bm{\alpha}_{0,K}^*}^{\top})),$$
$$\Gamma_f=\tilde{\sigma}^{-2}(\mathcal{G}_2^{-1})\mbox{E}(\bm{Z}_n^{\top}\bm{F}_0\bm{F}_0^{\top}\bm{Z}_n)\mathcal{G}_2^{-1}.$$ 
We then derive the following result.
\begin{theorem}
\label{thm5}
Under Assumption (A2), (A5) and (A6), it holds that
\begin{equation*}
\frac{N(\widehat{\bm{A}}^*-\bm{A}_0^*)^{\top}\Gamma_f(\widehat{\bm{A}}^*-\bm{A}_0^*)-KD}{\sqrt{2KD}}\overset{d}\to\mathcal{N}(0,1),\qquad \mbox{as} \ N\to\infty.
\end{equation*}
\end{theorem}
\section{Simulation Studies}
\label{sim}
\subsection{General Setting}
In this section, we study the finite-sample properties of the developed methodologies through numerical experiments. Since the intercept is not involved in group detection, we simplify the model by omitting the intercept. We simulate data from the following multiple functional regression model $y_n=\sum_{j=1}^p\langle X_{nj},\beta_j\rangle+\epsilon_n,\ n=1,\ldots,N,$ 
where $\epsilon_n\stackrel{i.i.d.}{\sim}\mathcal{N}(0,s^2)$. The functional covariates $\{X_{nj}(t)\colon j=1,\ldots,p,\ n\ge1\}$ and the coefficient functions $\{\beta_{j}(t)\colon j=1,\ldots,p\}$ are generated from the following basis expansion,
\begin{equation*}
\beta_{j}(t)=\sum_{d=1}^Db_{jd}\nu_d(t),\ \ X_{nj}(t)=\sum_{d=1}^D\xi_{nj,d}\nu_d(t),\ \ \xi_{nj,d}\sim\mathcal{N}(0,d^{-1.2}),
\end{equation*}
where $\{\nu_d(t)\colon d\ge1\}$ are Fourier basis functions. 
The coefficient scores are generated from three different templates:
\begin{itemize}
 \setlength\itemsep{-0.0em}
\item[$\delta_1$:] $\{b_{j1},\ldots,b_{jD}\}=f_j\times\{(D+1)/2,\ldots,2,1,2,\ldots,(D+1)/2\}$\hspace{2.22cm} (V-shape)
\item[$\delta_2$:] $\{b_{j1},\ldots,b_{jD}\}=f_j\times\{2^{-d}\colon d=1,\ldots,5\}$\hspace{5.35cm} (fast-decay)
\item[$\delta_3$:] $\{b_{j1},\ldots,b_{jD}\}=f_j\times\{1.2^{-d}\colon d=1,\ldots,5\}$\hspace{5.03cm} (slow-decay)
\end{itemize}
and $\{f_j\colon j=1,\ldots,10\}=(0.57, 0.75, 0.92, 5.20, 6.76, 8.32, 6.24, 2.17, 2.83, 3.48)$. In the simulation, we set $D=5$ and $p =10$. Results under different values of $D$ and $p$ are provided in the supplementary material. Table \ref{setting} displays the simulated coefficient scores. 
\begin{table}[ht]
\renewcommand{\arraystretch}{1}
\centering
\captionsetup{justification=centering}
	\caption{Coefficient scores $b_{jd}$ and the grouping structure.}
	\begin{tabular}{|c|ccc|cccc|ccc|}
	\hline 
        \multirow{2}{*}{\diagbox{$d$}{$j$}} &\multicolumn{3}{c|}{$\delta_1$} &\multicolumn{4}{c|}{$\delta_2$}&\multicolumn{3}{c|}{$\delta_3$}\\
        \cline{2-11}
           &1&2&3&4&5&6&7&8&9&10\\
           \hline 
        1 & 1.73& 2.25& 2.77& 2.60& 3.38& 4.15& 3.12&1.81 &2.35 &2.90\\
        2 & 1.15& 1.50& 1.84& 1.30& 1.69& 2.08& 1.56&1.50 &1.96 &2.41\\
        3 & 0.58& 0.75& 0.92& 0.65& 0.84& 1.04& 0.78&1.26 &1.63 &2.01\\
        4 & 1.15& 1.50& 1.84& 0.32& 0.42& 0.51& 0.39&1.05 &1.36 &1.68\\
        5 & 1.73& 2.25& 2.77& 0.16& 0.21& 0.26& 0.19&0.87 &1.13 &1.40\\
        \hline 
         \end{tabular}
	\label{setting}
\end{table}
Three concave functions $J_\lambda(\cdot)$ are considered for comparison: truncated LASSO (TLASSO) penalty function, MCP function, and SCAD penalty function. For a fair comparison, the same set of random seed is used across all penalty settings. Our objective is to demonstrate the influence of relevant parameters on group detection performance and to showcase the superiority of the newly proposed grouped model over existing regression models. In Section \ref{path}, we further investigate the grouping paths and discuss the necessity of using concave penalty functions  $J_\lambda(\cdot)$ in group detection. In Section \ref{tuning}, we study the performance of group detection in different settings. In Section \ref{pred}, we demonstrate the superiority of the detected grouped model in terms of prediction accuracy, compared to the ordinary multiple functional regression model and the matrix-variate regression model.
\subsection{Grouping Paths of Different Penalties}
\label{path}
\begin{figure}[ht]
\center
\includegraphics[scale=0.12]{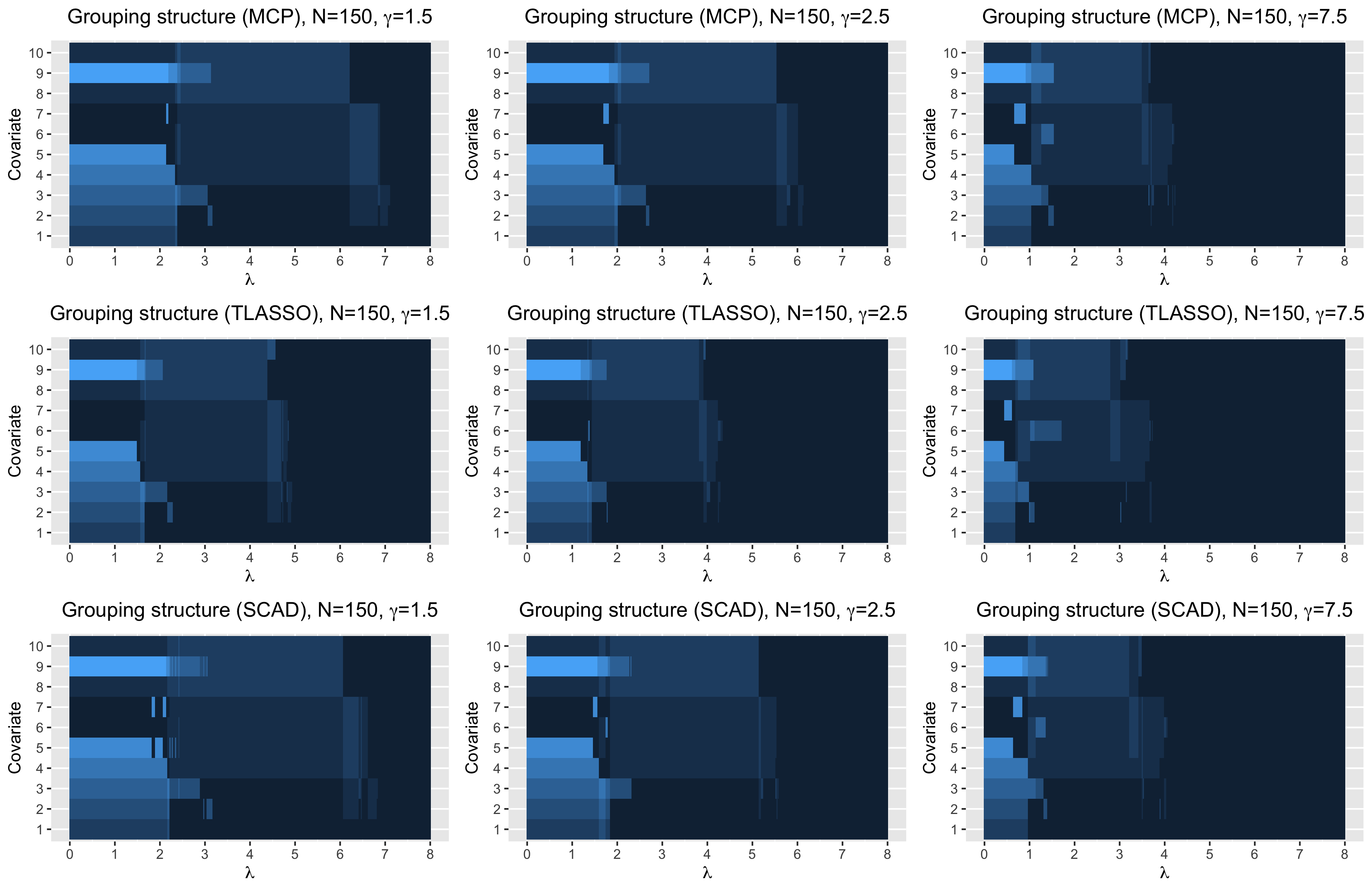}
\caption{Paths of grouping structure ($s=1.5$, $N=150$, $\tilde{\lambda}=0.15$). The x-axis represents the values of $\lambda$, and the y-axis represents the covariate indexes. }
\label{path-N150-s15}
\end{figure}

\begin{figure}[H]
\center
\includegraphics[scale=0.12]{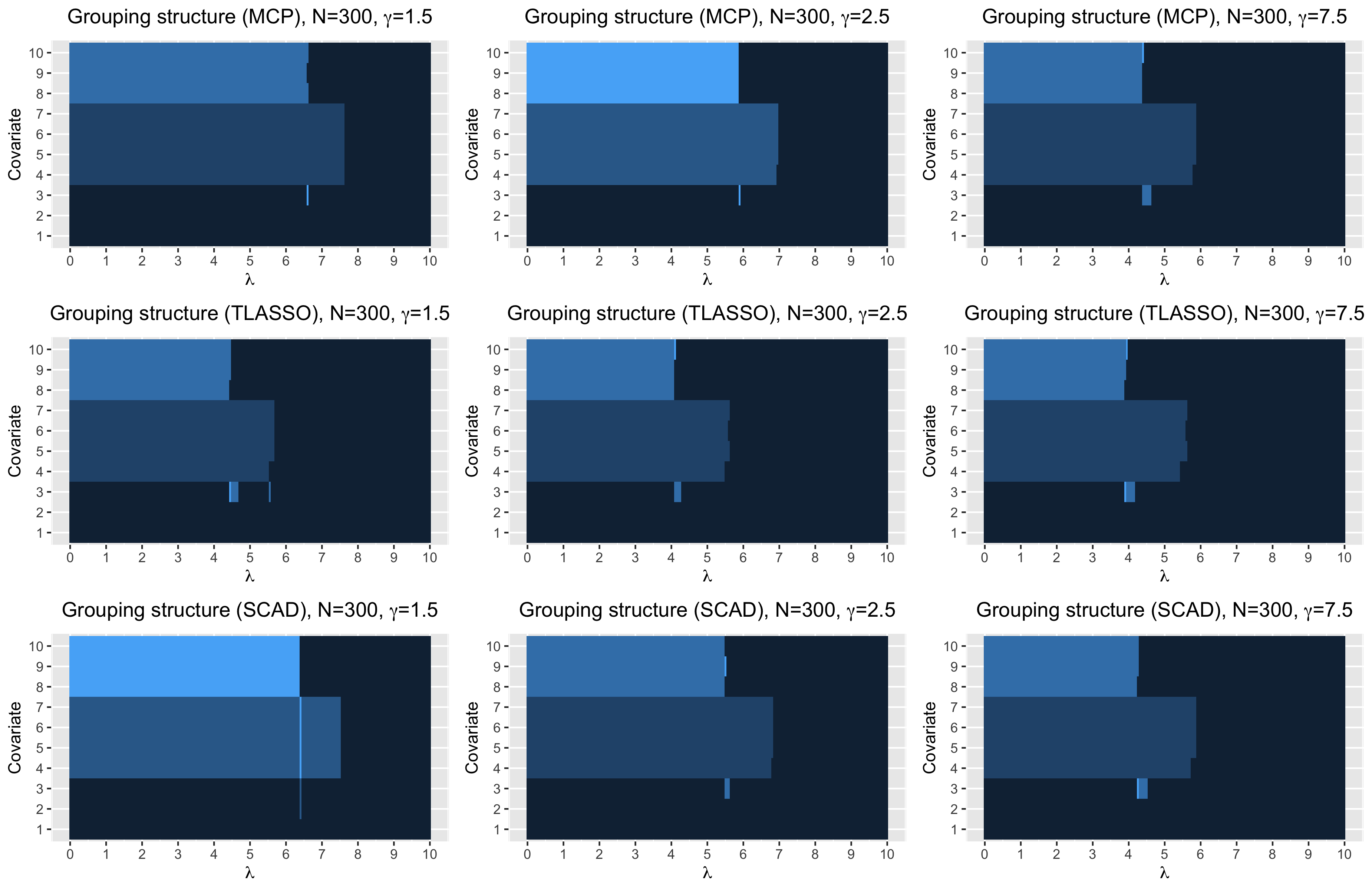}
\caption{Paths of grouping structure ($s=1.5$, $N=300$, $\tilde{\lambda}=0.15$). The x-axis represents the values of $\lambda$, and the y-axis represents the covariate indexes. }
\label{path-N300-s15}
\end{figure}

Figures \ref{path-N150-s15} and \ref{path-N300-s15} display the grouping paths against the value of $\lambda$ based on 150 and 300 samples, respectively. In these figures, covariates are grouped together at a specific value of $\lambda$ when they are marked in the same color. The x-axis represents the values of $\lambda$, and the y-axis corresponds to the covariate indexes. We set $\gamma=1.5$, 2.5, 7.5, $s=1.5$, and $\tilde{\lambda}=0.15$. 
As $\lambda$ increases, more covariates are grouped and eventually all are aggregated together when $\lambda$ is sufficiently large. 
The true grouping structure is detected over a wider range of $\lambda$ for the MCP and SCAD penalty. A comparison of the three columns in Figures \ref{path-N150-s15} and \ref{path-N300-s15} highlights the importance of using concave functions for $J_\lambda(\cdot)$. Notably, the grouping paths at $\gamma=7.5$ differ significantly from those at $\gamma=1.5, 2.5$. As $\gamma$ increases, the impact of the convexity of $J_\lambda(\cdot)$ diminishes, causing $J_\lambda(\cdot)$ to resemble the $\ell_1$-norm. Consequently, different covariates are aggregated at smaller values of $\lambda$. and the true grouping structure is identified over a narrower range of $\lambda$, making it more challenging to detect. It is well known that $\ell_1$-norm penalty leads to biased estimates and over-shrinkage. The constant part of $J_\lambda(\cdot)$ preserves the pronounced coefficient shape misalignment between groups, mitigating the over-shrinkage issue, and thereby significantly improves the chances of identifying the true grouping structure.

\subsection{Group Detection and Tuning Parameters}
\label{tuning}
In this section, we investigate the group detection performance of our regularization approach in various settings. For each scenario, we repeat the simulation 300 times, with each repetition involving $N=150$ or $N=300$ samples. A grid of $\lambda$ values is selected, and at the maximal $\lambda$ all covariates are grouped together. The Monte-Carlo cross-validation (MCCV) method (see \cite{picard1984cross}) is applied to determine the optimal grouping structure in each simulation run. After obtaining the grouping path over the grid of $\lambda$ values in each repetition, the $N$ samples are randomly split into the training set $\bm{\mathcal{X}}^b_{train}$ (with size $|\bm{\mathcal{X}}^b_{train}|=2N/3$) and the testing set $\bm{\mathcal{X}}^b_{test}$ (with size $|\bm{\mathcal{X}}^b_{test}|=N/3$), where $b$ represents the MCCV repetition. The training set is used to estimate the grouped models, which are then used for prediction on the testing set. We repeat this MCCV procedure 400 times, and the grouping structure associated with the minimum average prediction rooted mean square error (RMSE), defined as 
\begin{equation}
\label{RMSE}
\min_\lambda\frac{1}{400}\sum_{b=1}^{400}\sqrt{\frac{1}{|\bm{\mathcal{X}}^b_{test}|}\sum_{n\in\bm{\mathcal{X}}^b_{test}}(\hat{y}^{\lambda}_n-y_n)^2}
\end{equation}
is selected, where $\hat{y}_n^{\lambda}$ represents the predicted value of $y_n$ obtained from the grouped model detected under $\lambda$. 

\begin{figure}[ht]
\center
\includegraphics[scale=0.15]{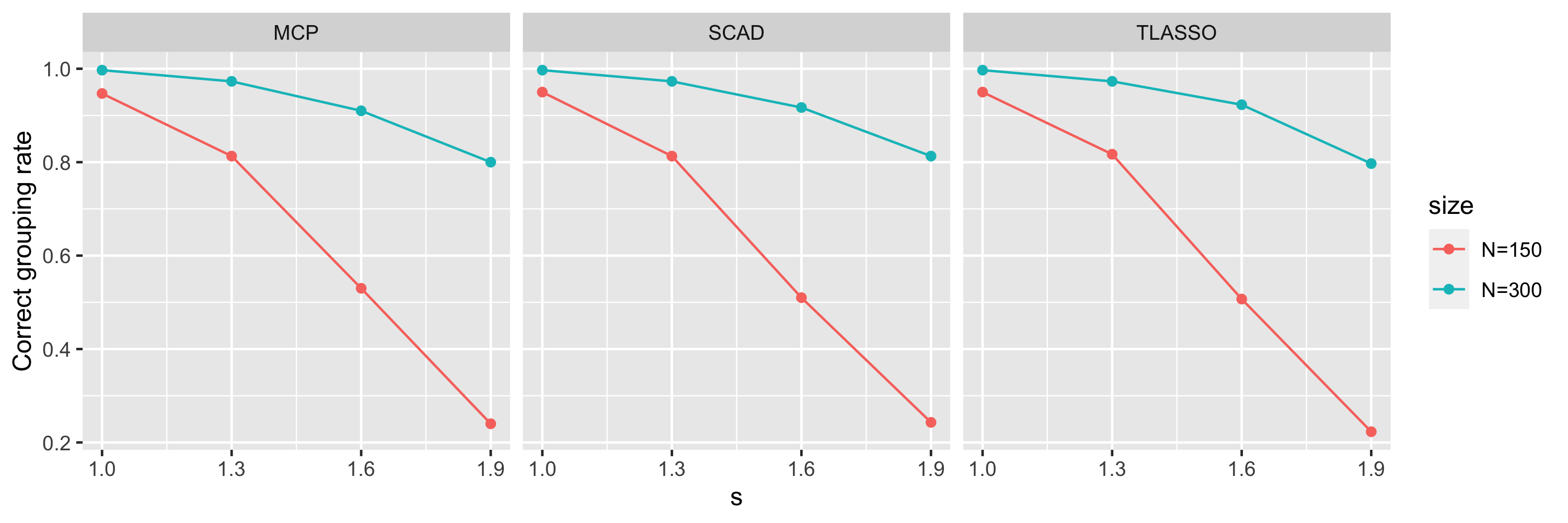}
\caption{Average correct grouping rate over all the simulation runs ($\tilde{\lambda}=0.2$).}
\label{group_var}
\end{figure}

To assess the impact of the signal-to-noise ratio on group detection performance, we generate samples with different variance $s^2$ of the random errors $\{\epsilon_n\colon n\ge1\}$, and set $\gamma=2.1$, $\tilde{\lambda}=0.2$. We compare the performance of the three penalty functions by calculating the proportion of simulation runs where the true grouping structure is correctly detected (correct grouping rate), as shown in Figure \ref{group_var}. The results indicate that, as $s$ increases, correctly grouping covariates becomes more challenging. This is because the increased data variation diminishes the distinctiveness of coefficient shape homogeneity. When the sample size increases from 150 to 300, the ability of detecting the correct grouping structure is substantially improved. The three penalty functions perform similarly and achieve nearly-perfect group detection when $N=300$ and $s=1$. 

\begin{figure}[ht]
\center
\includegraphics[scale=0.11]{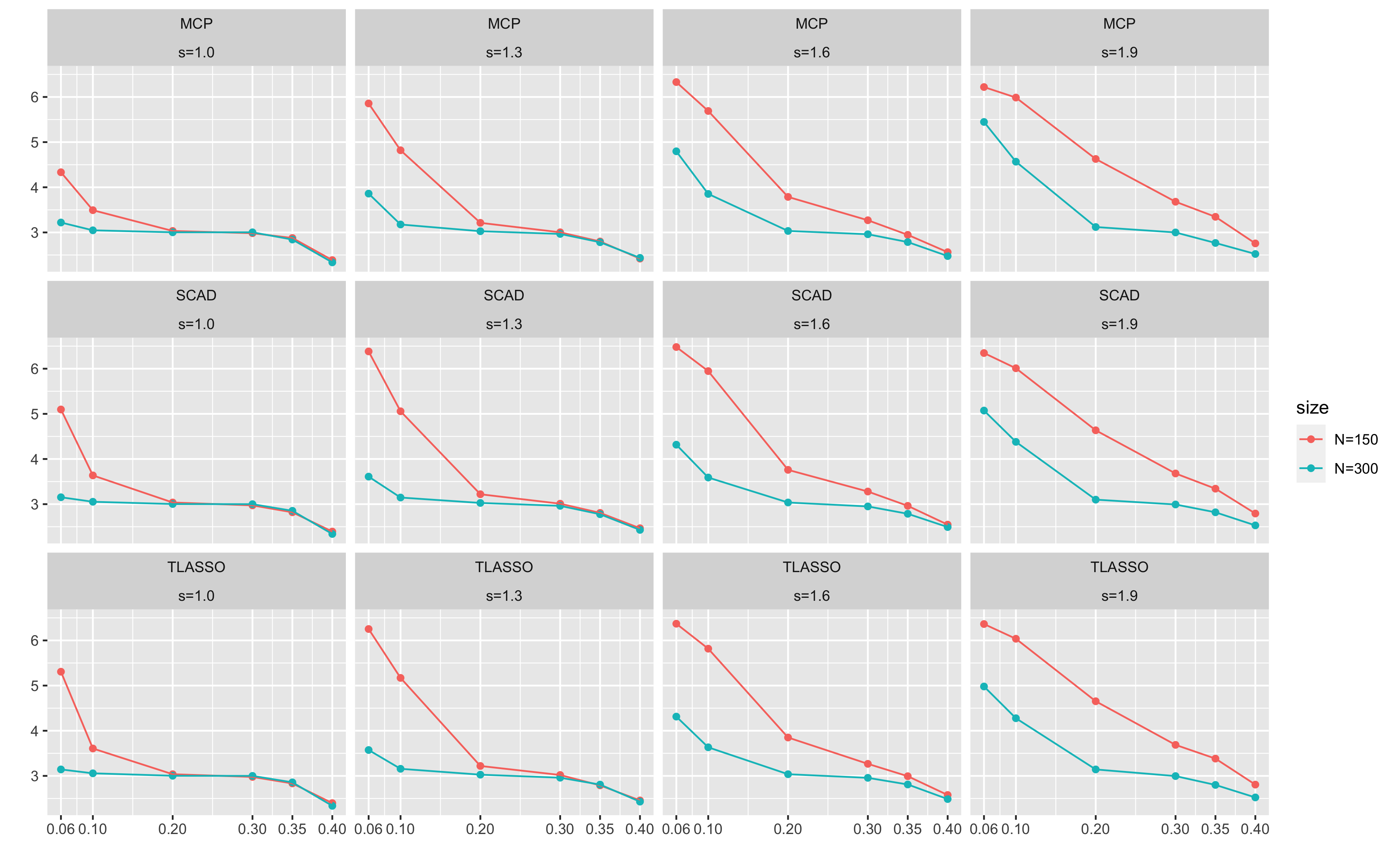}
\caption{Average number of groups in the detected grouping structure over all the simulation runs. In each figure, the $x$-axis represents the thresholds $\tilde{\lambda}$, and the $y$-axis represents the average number of groups in the detected grouping structure.}
\label{gnumber}
\end{figure}

To examine the effect of the threshold $\tilde{\lambda}$ on group detection, we set $\tilde{\lambda}=$0.06, 0.1, 0.2, 0.3, 0.35, 0.4. Generally, a larger threshold results in a more parsimonious grouping structure, meaning fewer groups are formed. However, if $\tilde{\lambda}$ is too large, all covariates may be aggregated into a single group, leading to model misspecification and suboptimal predictive performance. Our goal is to determine whether group detection performance remains robust across different selections of $\tilde{\lambda}$. Figure \ref{gnumber} illustrates the average number of groups in the detected grouping structure. We observe that the results are indeed robust to the choice of threshold, particularly when the sample size is large or the signal-to-noise ratio is high.

Figure \ref{rate} displays the average correct grouping rate over all  simulation runs. Figures 1 in the supplementary material displays the average prediction RMSE of the detected grouped model. Notably, when $\tilde{\lambda}$ is selected around $0.2$, we achieve the best group detection performance across all settings, and the associated grouped model yields the highest predictive accuracy. Conversely, when over-shrinkage occurs, due to overly large values of $\tilde{\lambda}$, the prediction accuracy can significantly decline due to model misspecification. Furthermore, when covariates within the same group are not aggregated effectively, prediction performance suffers due to increased estimation error.

\begin{figure}[ht]
\center
\includegraphics[scale=0.11]{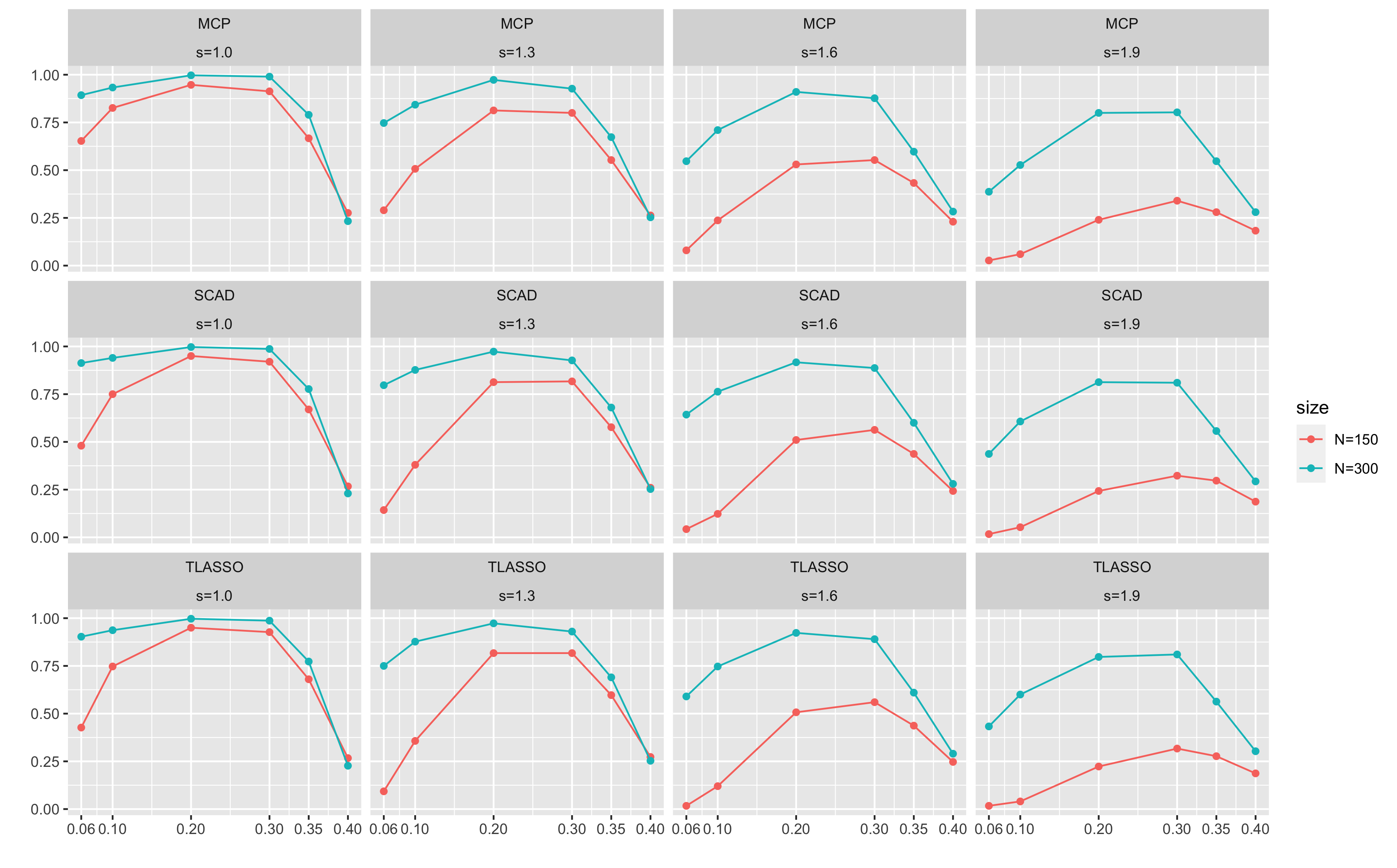}
\caption{Average correct grouping rate over all the simulation runs. The $x$-axis represents the thresholds $\tilde{\lambda}$, and the $y$-axis represents the average correct grouping rate.}
\label{rate}
\end{figure}

\subsection{Prediction Performance}
\label{pred}
In this section, we compare the prediction performance of our grouped model (Grouped) with the other two competing regression models, including the ordinary multiple functional regression model (Ordinary) and the matrix-variate regression model (Matrix). In the matrix-variate regression model, $(\bm{\xi}_{n1}|\ldots|\bm{\xi}_{np})^{\top}$ is treated as the covariate matrix. For our grouped model, we use the MCP function as $J_\lambda(\cdot)$ to detect the grouping structure. Additionally, we consider a grouped regression model based on the oracle grouping structure (Oracle) to assess the impact of uncertainty in group detection on predictive power.

\begin{figure}[ht]
\center
\includegraphics[scale=0.12]{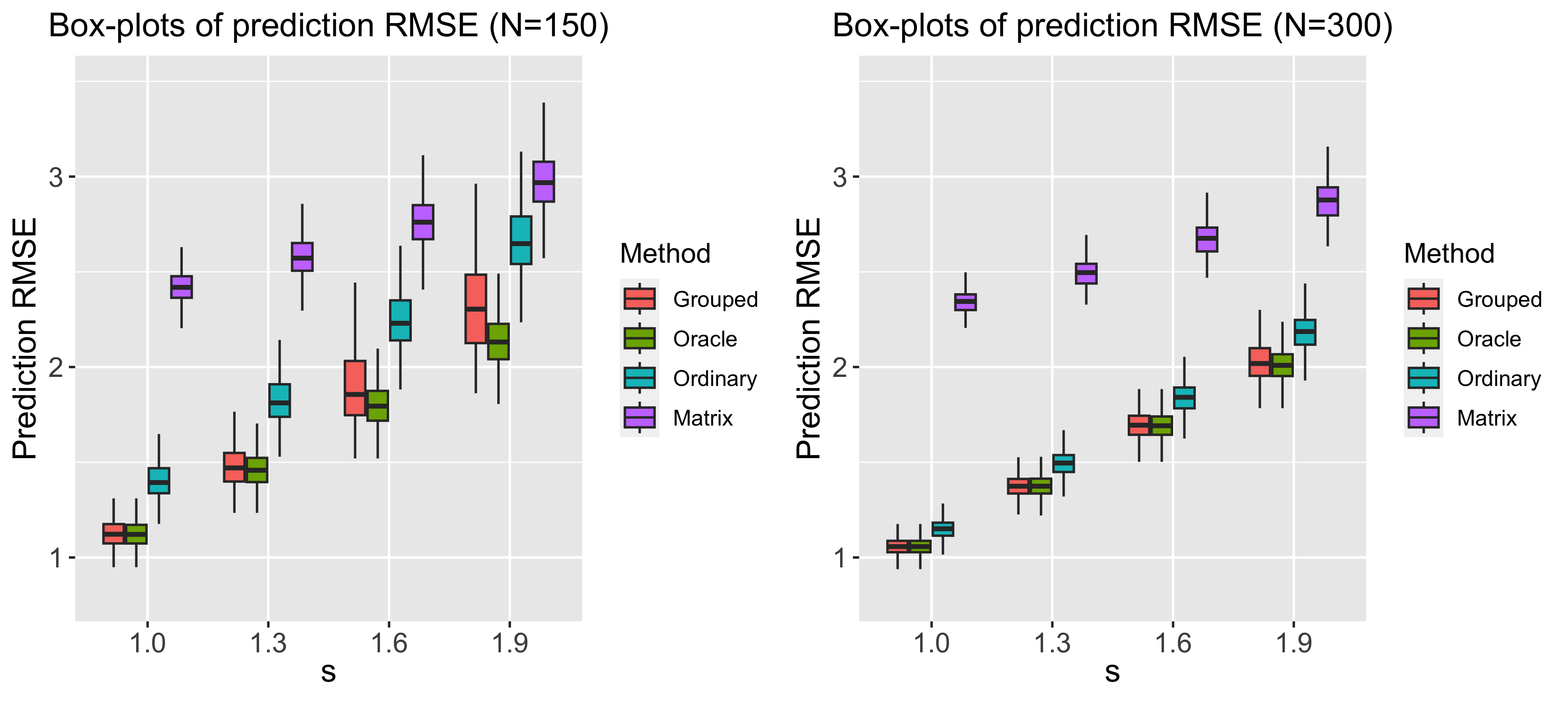}
\caption{Box-plots of the prediction RMSE of the four methods.}
\label{predbox}
\end{figure}

The prediction errors are obtained from 300 simulation runs, applying the same Monte Carlo cross-validation (MCCV) procedure to calculate the prediction RMSE. Across all settings, our grouped model consistently produces the best predictions, while the matrix-variate regression model performs the worst. This indicates that when covariates with substantial heterogeneity are modeled using a single template coefficient, the bias from model misspecification can be significant. Therefore, it is crucial to properly segment the covariates into distinct groups and assign different template coefficients accordingly. As the variation of the random error increases, the resulting estimation error can diminish predictive power. However, our detected grouped model still outperforms the other models in terms of prediction accuracy.

The predictive superiority of our grouped model is expected, as both the ordinary multiple functional regression model and the matrix-variate regression model fall within our grouped regression framework. Specifically, the matrix-variate model can be seen as a grouped model in which all covariates are combined into a single group, while the ordinary multivariate regression model can be viewed as a special case where no covariates are grouped. Thus, the grouped model based on the detected grouping structure effectively balances variance and bias, consistently delivering predictions that are at least as good as those obtained from these two alternative models.
 
\section{Real Data Analysis}
\label{rd}
In sugar manufacturing, it is crucial to analyze and monitor the fluorescence spectra of sugar samples, as these are closely related to sugar purity. By utilizing spectrofluorometry and chemometrics, the beet sugar manufacturing process can be better understood (see \cite{munck1998chemometrics} and \cite{bro1999exploratory}). In this study, we apply our new method to the dataset used in \cite{munck1998chemometrics} and \cite{bro1999exploratory} to assess its practical performance.

\begin{figure}[ht]
\center
\includegraphics[scale=0.16]{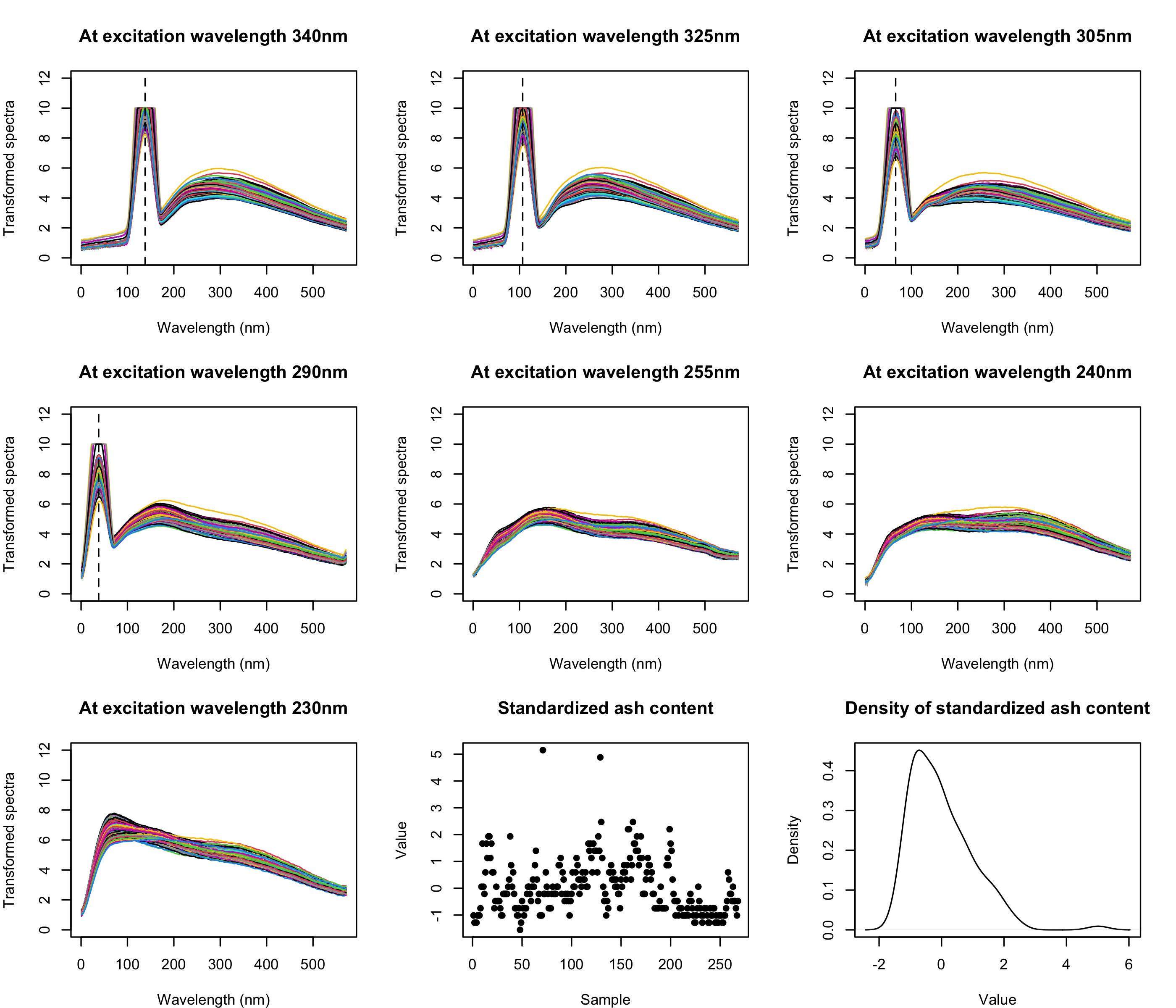}
\caption{Spectra curves evaluated at seven excitement wavelengths,  standardized ash content, and the density function of standardized ash content.}
\label{f1}
\end{figure}

The dataset consists of 268 sugar samples collected during the three months of operation in late autumn from a sugar plant in Scandinavia. For each sample, emission spectra from 275 to 560 were recorded at 0.5nm intervals, resulting in 571 observations for each spectra curve. The emission spectra were measured at seven excitement wavelengths (340nm, 325nm, 305nm, 290nm, 255nm, 240nm, 230nm), as shown in Figure \ref{f1}. At the first four excitement wavelength (340nm, 325nm, 305nm, 290nm), the emission spectra curves exhibit a peak, which shifts leftward as the wavelength decreases and eventually disappears. In addition to emission spectra, the dataset includes other purity-related measurements, such as ash content, which is a key indicator of sugar quality. Our goal is to explore the relationship between ash content and the emission spectra and examine the homogeneity of spectra curves recorded at different excitation wavelengths. By grouping the excitation wavelengths, food scientists can identify common patterns in the relationship between sugar quality and the spectra, develop better evaluation models, and improve assessment efficiency.

In the analysis, the seven emission spectra curves for each sample serve as the functional covariates, while the standardized ash contents serves as the response variable. In this application, $X_1(t)$ represents the spectra curves measured at an excitation wavelength of 340 nm, $X_2(t)$ at 325nm, and so on. MCCV is used to select the grouping structure. To stabilize variance, we apply a cubic root transformation to the spectra curves. All the spectra curves are pooled together to calculate the covariance operator, and the associated eigenfunctions are employed as the basis functions for dimension reduction. The number of basis functions $D$ is selected such that the cumulative proportion of the associated eigenvalues exceeds 90\%, which results in four basis functions. 

In the selected grouping structure, the excitement wavelengths 340nm, 325nm, 305nm,  and 290nm are clustered together, yielding the following groups
$\hat{\delta}_1\colon$340nm, 325nm, 305nm, 290nm; 
$\hat{\delta}_2\colon$255nm; 
$\hat{\delta}_3\colon$240nm; 
$\hat{\delta}_4\colon$230nm. 
This result suggests that the peak location in the spectra curves for excitation wavelengths of 340 nm, 325 nm, 305 nm, and 290 nm is not strongly related to the association between ash content and emission spectra. However, since these four wavelengths are grouped separately from the others, the peak magnitude must be considered when analyzing this relationship.

\begin{figure}[ht]
\center
\includegraphics[scale=0.13]{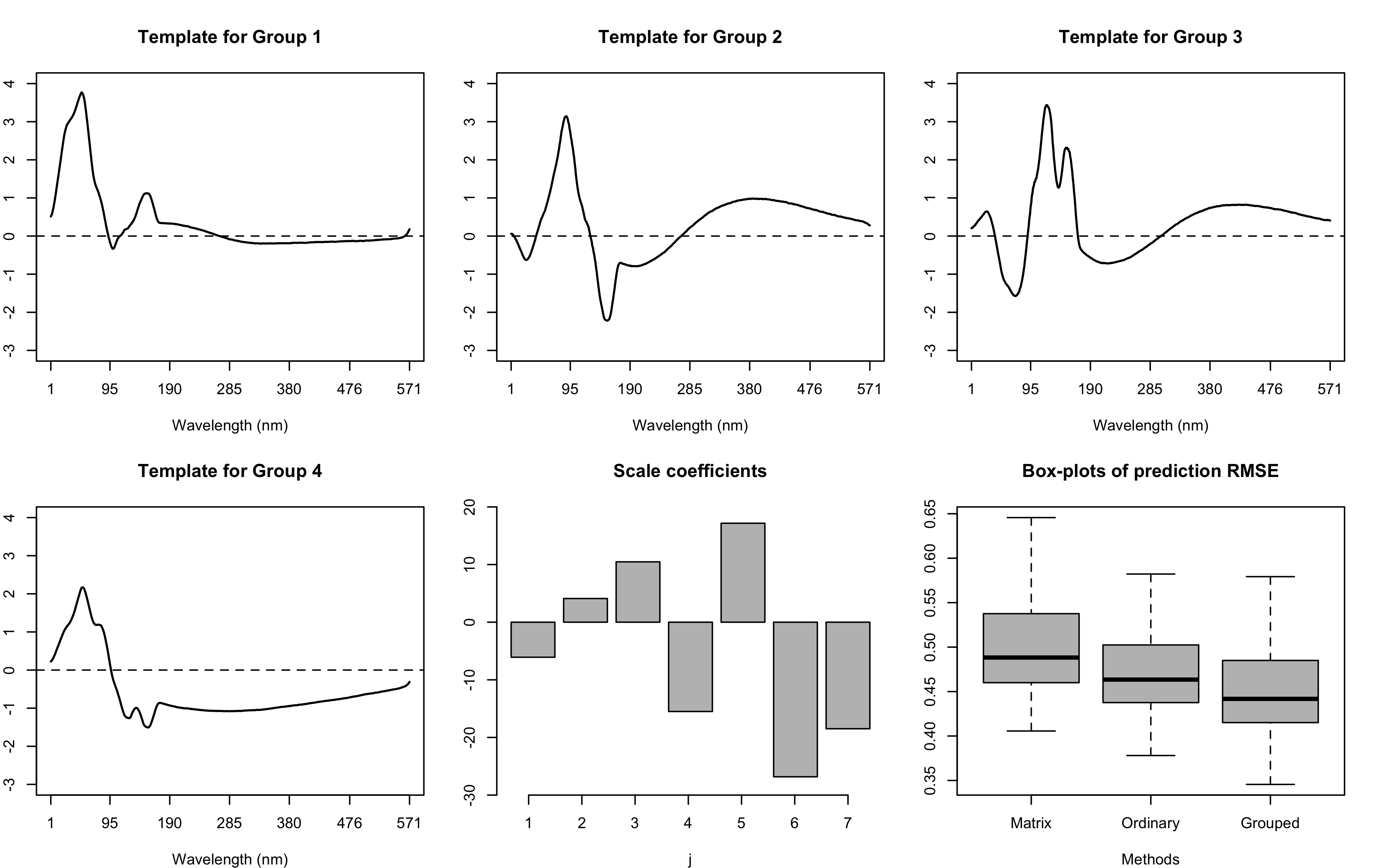}
\caption{Normalized template coefficient functions for different groups, the associated scale coefficients, and the box-plots of the prediction RMSE of the three competing methods.}
\label{f2}
\end{figure}

To evaluate the prediction performance of the grouped model, we compare the prediction root mean square error (RMSE) of ash content from the three regression models used in our simulation study. The normalized estimated template coefficient functions, the associated scale coefficients, and the box plots of prediction RMSE of the three competing methods are shown in Figure \ref{f2}. From the estimated template coefficient functions, we find that the emission spectra curves over the range [0,190nm] have a stronger association with ash content. The estimated scale coefficients indicate that the spectra obtained at excitation wavelengths of 240nm and 230nm are more closely related to ash content. The box plots of prediction RMSE demonstrate that the grouped model provides the best explanation of the underlying data mechanisms.
\section{Conclusion}
\label{con}
In this article, we introduce a novel grouped multiple functional regression framework based on coefficient shape homogeneity. The modeling procedure involves two main steps: first, detecting the underlying grouping structure; and second, developing the proposed grouped multiple functional regression model based on the detected grouping structure. The novelty of this work lies in several key contributions: 1.) We introduce a novel grouped multiple functional regression model based on the homogeneity of coefficient shape, rather than relying on coefficient equality as in the existing literature. This makes our regression framework more inclusive and general; 2.) We propose a new ``shape-alignment" regularization approach that incorporates a novel pairwise coefficient shape misalignment penalty to detect the unknown grouping structure. We thoroughly examine the consistency properties of the detected grouping structure and the asymptotic properties of the model estimates. The entire procedure is fully data-driven, making it applicable to a wide range of cases and offering a new, parsimonious modeling strategy for complex multivariate functional data.

We identify several research topics for future work. The regularization method developed here has the potential to enhance other functional data methodologies, such as functional data clustering, functional factor models, and non-linear multiple functional regression with deep neural networks. These new regularization techniques can be uniformly expressed in the form of a loss function combined with a pairwise coefficient shape misalignment penalty. However, extending this approach is not straightforward and will require significant additional work to develop these methodologies. In terms of application, the methods developed in this paper have the potential to benefit numerous fields, including but not limited to neuroimaging analysis, meteorological analysis, and traffic volume management. In neuroimaging, our methods could be used to unravel brain functional connectivity, as highly connected brain subregions often produce synchronized signals, allowing for the use of common-template coefficient functions when treating brain signals at each electrode as covariates. In meteorological analysis, grouping analysis can reveal interdependence among climate features across different locations and contribute to more accurate joint forecasts of future climate dynamics. In traffic volume management, our methods provide a means to study the persistence of traffic flow patterns across different roads in a city, facilitating better joint prediction of traffic volumes, improving level-of-service determination, and reducing congestion. These application studies will be pursued in future research.

\bibliography{Grouping-arkiv}

@article{bro1999exploratory,
  title={Exploratory study of sugar production using fluorescence spectroscopy and multi-way analysis},
  author={Bro, Rasmus},
  journal={Chemometrics and Intelligent Laboratory Systems},
  volume={46},
  number={2},
  pages={133--147},
  year={1999},
  publisher={Elsevier}
}

@article{cai2006prediction,
  title={Prediction in functional linear regression},
  author={Cai, T Tony and Hall, Peter},
   journal={The Annals of Statistics},
   volume={34},
  number={5},
  pages={2159--2179},
  year={2006}
}

@article{she2022supervised,
  title={Supervised multivariate learning with simultaneous feature auto-grouping and dimension reduction},
  author={She, Yiyuan and Shen, Jiahui and Zhang, Chao},
  journal={Journal of the Royal Statistical Society Series B: Statistical Methodology},
  volume={84},
  number={3},
  pages={912--932},
  year={2022},
  publisher={Oxford University Press}
}

@book{hall2014martingale,
  title={Martingale limit theory and its application},
  author={Hall, Peter and Heyde, Christopher C},
  year={2014},
  publisher={Academic press}
}

@article{ghorai1980asymptotic,
  title={Asymptotic normality of a quadratic measure of orthogonal series type density estimate},
  author={Ghorai, Jugal},
  journal={Annals of the Institute of Statistical Mathematics},
  volume={32},
  pages={341--340},
  year={1980},
  publisher={Elsevier}
}

@article{chiou2016multivariate,
  title={Multivariate functional linear regression and prediction},
  author={Chiou, Jeng-Min and Yang, Ya-Fang and Chen, Yu-Ting},
  journal={Journal of Multivariate Analysis},
  volume={146},
  pages={301--312},
  year={2016},
  publisher={Elsevier}
}

@article{ke2015homogeneity,
  title={Homogeneity pursuit},
  author={Ke, Zheng Tracy and Fan, Jianqing and Wu, Yichao},
  journal={Journal of the American Statistical Association},
  volume={110},
  number={509},
  pages={175--194},
  year={2015},
  publisher={Taylor \& Francis}
}

@article{mahzarnia2022multivariate,
  title={Multivariate functional group sparse regression: Functional predictor selection},
  author={Mahzarnia, Ali and Song, Jun},
  journal={PloS one},
  volume={17},
  number={4},
  pages={e0265940},
  year={2022},
  publisher={Public Library of Science San Francisco, CA USA}
}

@article{tibshirani1996regression,
  title={Regression shrinkage and selection via the lasso},
  author={Tibshirani, Robert},
  journal={Journal of the Royal Statistical Society Series B: Statistical Methodology},
  volume={58},
  number={1},
  pages={267--288},
  year={1996},
  publisher={Oxford University Press}
}

@article{zhang2010nearly,
  title={Nearly unbiased variable selection under minimax concave penalty},
  author={Zhang, Cun-Hui},
  journal={The Annals of Statistics},
  volume={38},
  number={2},
  pages={894--942},
  year={2010}
}

@article{zhou2013tensor,
  title={Tensor regression with applications in neuroimaging data analysis},
  author={Zhou, Hua and Li, Lexin and Zhu, Hongtu},
  journal={Journal of the American Statistical Association},
  volume={108},
  number={502},
  pages={540--552},
  year={2013},
  publisher={Taylor \& Francis}
}

@article{fan2001variable,
  title={Variable selection via nonconcave penalized likelihood and its oracle properties},
  author={Fan, Jianqing and Li, Runze},
  journal={Journal of the American statistical Association},
  volume={96},
  number={456},
  pages={1348--1360},
  year={2001},
  publisher={Taylor \& Francis}
}

@article{james2009functional,
  title={Functional linear regression that’s interpretable},
  author={James, Gareth M and Wang, Jing and Zhu, Ji},
   journal={The Annals of Statistics},
  volume={37},
  number={5},
  pages={2083--2108},
  year={2009}
}

@article{lian2013shrinkage,
  title={Shrinkage estimation and selection for multiple functional regression},
  author={Lian, Heng},
  journal={Statistica Sinica},
  volume={23},
  number={1},
  pages={51--74},
  year={2013},
  publisher={JSTOR}
}

@article{munck1998chemometrics,
  title={Chemometrics in food science—a demonstration of the feasibility of a highly exploratory, inductive evaluation strategy of fundamental scientific significance},
  author={Munck, L and N{\o}rgaard, L and Engelsen, S Balling and Bro, R and Andersson, CA},
  journal={Chemometrics and Intelligent Laboratory Systems},
  volume={44},
  number={1-2},
  pages={31--60},
  year={1998},
  publisher={Elsevier}
}

@article{yeh2023variable,
  title={Variable Selection in Multivariate Functional Linear Regression},
  author={Yeh, Chi-Kuang and Sang, Peijun},
  journal={Statistics in Biosciences},
  pages={1--18},
  year={2023},
  publisher={Springer}
}

@article{zhao2012wavelet,
  title={Wavelet-based LASSO in functional linear regression},
  author={Zhao, Yihong and Ogden, R Todd and Reiss, Philip T},
  journal={Journal of Computational and Graphical Statistics},
  volume={21},
  number={3},
  pages={600--617},
  year={2012},
  publisher={Taylor \& Francis}
}

@article{picard1984cross,
  title={Cross-validation of regression models},
  author={Picard, Richard R and Cook, R Dennis},
  journal={Journal of the American Statistical Association},
  volume={79},
  number={387},
  pages={575--583},
  year={1984},
  publisher={Taylor \& Francis}
}

@article{yuan2006model,
  title={Model selection and estimation in regression with grouped variables},
  author={Yuan, Ming and Lin, Yi},
  journal={Journal of the Royal Statistical Society Series B: Statistical Methodology},
  volume={68},
  number={1},
  pages={49--67},
  year={2006},
  publisher={Oxford University Press}
}

@article{tibshirani2005sparsity,
  title={Sparsity and smoothness via the fused lasso},
  author={Tibshirani, Robert and Saunders, Michael and Rosset, Saharon and Zhu, Ji and Knight, Keith},
  journal={Journal of the Royal Statistical Society Series B: Statistical Methodology},
  volume={67},
  number={1},
  pages={91--108},
  year={2005},
  publisher={Oxford University Press}
}

@article{wang2016functional,
  title={Functional data analysis},
  author={Wang, Jane-Ling and Chiou, Jeng-Min and M{\"u}ller, Hans-Georg},
  journal={Annual Review of Statistics and its application},
  volume={3},
  pages={257--295},
  year={2016},
  publisher={Annual Reviews}
}

@article{lin2017locally,
  title={Locally sparse estimator for functional linear regression models},
  author={Lin, Zhenhua and Cao, Jiguo and Wang, Liangliang and Wang, Haonan},
  journal={Journal of Computational and Graphical Statistics},
  volume={26},
  number={2},
  pages={306--318},
  year={2017},
  publisher={Taylor \& Francis}
}

@article{gabay1976dual,
  title={A dual algorithm for the solution of nonlinear variational problems via finite element approximation},
  author={Gabay, Daniel and Mercier, Bertrand},
  journal={Computers \& mathematics with applications},
  volume={2},
  number={1},
  pages={17--40},
  year={1976},
  publisher={Elsevier}
}

@article{glowinski1975approximation,
  title={Sur l'approximation, par {\'e}l{\'e}ments finis d'ordre un, et la r{\'e}solution, par p{\'e}nalisation-dualit{\'e} d'une classe de probl{\`e}mes de Dirichlet non lin{\'e}aires},
  author={Glowinski, Roland and Marroco, Americo},
  journal={Revue fran{\c{c}}aise d'automatique, informatique, recherche op{\'e}rationnelle. Analyse num{\'e}rique},
  volume={9},
  number={R2},
  pages={41--76},
  year={1975},
  publisher={EDP Sciences}
}

@book{ramsay2005functional,
  title={Functional Data Analysis (2nd ed.)},
  author={Ramsay, J.\ O. and Silverman, B.\ W.},
  journal={Springer Series in Statistics},
  year={2005},
  publisher={Springer, New York}.
}

@book{lazar2010numerical,
  title={Numerical Analysis for Statisticians (2nd ed.)},
  author={Lazar, K.},
  journal={Statistics and Computing},
  year={2010},
  publisher={Springer, New York}.
}

@article{ding2018matrix,
  title={Matrix variate regressions and envelope models},
  author={Ding, Shanshan and Dennis Cook, R},
  journal={Journal of the Royal Statistical Society Series B: Statistical Methodology},
  volume={80},
  number={2},
  pages={387--408},
  year={2018},
  publisher={Oxford University Press}
}

@article{hung2013matrix,
  title={Matrix variate logistic regression model with application to EEG data},
  author={Hung, Hung and Wang, Chen-Chien},
  journal={Biostatistics},
  volume={14},
  number={1},
  pages={189--202},
  year={2013},
  publisher={Oxford University Press}
}

@article{li2010dimension,
  title={On dimension folding of matrix-or array-valued statistical objects},
  author={Li, Bing and Kim, Min Kyung and Altman, Naomi},
  journal={The Annals of Statistics},
  volume={38},
  number={2},
  pages={1094--1121},
  year={2010}
}

@article{shen2010grouping,
  title={Grouping pursuit through a regularization solution surface},
  author={Shen, Xiaotong and Huang, Hsin-Cheng},
  journal={Journal of the American Statistical Association},
  volume={105},
  number={490},
  pages={727--739},
  year={2010},
  publisher={Taylor \& Francis}
}

@article{ma2017concave,
  title={A concave pairwise fusion approach to subgroup analysis},
  author={Ma, Shujie and Huang, Jian},
  journal={Journal of the American Statistical Association},
  volume={112},
  number={517},
  pages={410--423},
  year={2017},
  publisher={Taylor \& Francis}
}

@article{benning2015preconditioned,
  title={Preconditioned ADMM with nonlinear operator constraint},
  author={Benning, Martin and Knoll, Florian and Sch{\"o}nlieb, Carola-Bibiane and Valkonen, Tuomo},
  journal={IFIP Advances in Information and Communication Technology},
  volume={494},
  pages={117--126},
  year={2015},
  publisher={Springer}
}

@article{latorre2019fast,
  title={Fast and provable ADMM for learning with generative priors},
  author={Latorre, Fabian and Cevher, Volkan and others},
  journal={Advances in Neural Information Processing Systems},
  volume={32},
  year={2019}
}

@article{muller2005generalized,
  title={Generalized functional linear models},
  author={M{\"u}ller, Hans-Georg and Stadtm{\"u}ller, Ulrich},
  journal={The Annals of Statistics},
  volume={33},
   number={2},
   pages={774--805},
  year={2005}
}

@article{hall2006properties,
  title={On properties of functional principal components analysis},
  author={Hall, Peter and Hosseini-Nasab, Mohammad},
  journal={Journal of the Royal Statistical Society Series B: Statistical Methodology},
  volume={68},
  number={1},
  pages={109--126},
  year={2006},
  publisher={Oxford University Press}
}

@article{hall2007methodology,
  title={Methodology and convergence rates for functional linear regression},
  author={Hall, Peter and Horowitz, Joel L},
  journal={The Annals of Statistics},
  volume={35},
   number={1},
   pages={70--91},
  year={2007}
}

@article{jiao2023functional,
  title={Functional time series prediction under partial observation of the future curve},
  author={Jiao, Shuhao and Aue, Alexander and Ombao, Hernando},
  journal={Journal of the American Statistical Association},
  volume={118},
  number={541},
  pages={315--326},
  year={2023},
  publisher={Taylor \& Francis}
}
\bibliographystyle{agsm}

\end{document}